\documentclass{article}
\usepackage{xcolor}
\usepackage{url}  
\usepackage{natbib}  
\usepackage{fullpage}
\usepackage[english]{babel}
\usepackage{eurosym}
\usepackage{setspace}
\usepackage [autostyle, english = american]{csquotes} 
\MakeOuterQuote{"} 
\usepackage{booktabs}       
\usepackage{amsfonts}       
\usepackage{nicefrac}       
\usepackage{lipsum}
\usepackage{graphicx}
\usepackage{amsmath}
\usepackage{float}
\usepackage{soul}

\usepackage{xcolor}
\definecolor{mycyan}{RGB}{237, 252, 252} 
\usepackage[most]{tcolorbox}
\newtcolorbox{abstractbox}{
  enhanced,
  colframe=mycyan,
  colback=mycyan,
  opacityback=0.15,   
  opacityframe=1,
  boxrule=1pt,
  arc=2mm,
  left=8pt,right=8pt,top=8pt,bottom=8pt
}
\renewenvironment{abstract}
  {\begin{abstractbox}\noindent\textbf{Abstract.} }
  {\end{abstractbox}}

\title{\textbf{GENESIS: A Generative Model of Episodic–Semantic Interaction}}

\author{Marco D'Alessandro, Leo D'Amato, Mikel Elkano, Mikel Uriz, Giovanni Pezzulo$^{*}$  \\
\\
Institute of Cognitive Sciences and Technologies, National Research Council, Rome, Italy  \\
$*$ Corresponding author: giovanni.pezzulo@istc.cnr.it}

\begin{document}
\maketitle

\begin{abstract}
A central challenge in cognitive neuroscience is to explain how semantic and episodic memory—two major forms of declarative memory, typically associated with cortical and hippocampal processing—interact to support learning, recall, and imagination. Despite significant advances, we still lack a unified computational framework that jointly accounts for core empirical phenomena across both semantic and episodic processing domains. Here, we introduce the Generative Episodic–Semantic Integration System (GENESIS), a computational model that formalizes memory as the interaction between two limited-capacity generative systems: a Cortical-VAE, supporting semantic learning and generalization, and a Hippocampal-VAE, supporting episodic encoding and retrieval within a retrieval-augmented generation (RAG) architecture. GENESIS reproduces hallmark behavioral findings—including generalization in semantic memory, recognition, serial recall effects and gist-based distortions in episodic memory, and constructive episodic simulation—while capturing their dynamic interactions. The model elucidates how capacity constraints shape the fidelity and memorability of experiences, how semantic processing introduces systematic distortions in episodic recall, and how episodic replay can recombine previous experiences. Together, these results provide a principled account of memory as an active, constructive, and resource-bounded process. GENESIS thus advances a unified theoretical framework that bridges semantic and episodic memory, offering new insights into the generative foundations of human cognition.

\end{abstract}


\section{Introduction}

A key distinction in psychology and neuroscience is between semantic and episodic memory, the two major forms of declarative memory. Semantic memory refers to structured knowledge about facts and concepts, primarily supported by cortical systems, whereas episodic memory involves personally experienced events embedded in specific spatial and temporal contexts, mainly associated with the hippocampal formation \citep{tulving1972episodic,tulving2002episodic,knowlton1995remembering,eichenbaum1999hippocampus}.

Despite this distinction, a long-held assumption is that semantic and episodic systems play complementary roles, as formalized in the influential complementary learning systems (CLS) theory \citep{mcclelland1995there,norman2003modeling}. The CLS theory posits that experiences are rapidly encoded in the hippocampus and later replayed to gradually train cortical semantic representations. Debate persists, however, over whether episodic traces remain permanently stored in the hippocampus after cortical consolidation. Recent proposals extend CLS theory while maintaining its central tenets. These include the ideas that hippocampal replay is goal-directed, supports statistical learning, and that cortical learning can sometimes be rapid rather than incremental \citep{kumaran2016learning,kumaran2012generalization,schapiro2017complementary}. Another model that builds upon CLS introduces a Hopfield network for fast episodic encoding, coupled with a generative cortical model for semantic inference, allowing interactions such as pattern completion \citep{spens2024generative}. Finally, a few models move beyond the standard assumptions of CLS. One focuses on the generative aspects of episodic memory and on semantic completion, whereby the hippocampus retrieves partial cues and cortical networks fill in missing semantic details \citep{fayyaz2022model}. Another theoretical proposal suggests that semantic memory engages in statistical learning, while episodic memory plays the complementary role of preserving surprising experiences in a relatively raw format \citep{nagy2025adaptive}.

Despite these advances, major challenges remain. Most existing models assume that episodic memories are formed independently of semantic memories, by implementing them within two separate networks (but see \citep{fayyaz2022model,nagy2025adaptive,spens2025hippocampo}). However, several converging lines of evidence suggest a more integrated picture. First, the hippocampus receives highly processed inputs from the entorhinal cortex, implying that episodic encoding inherently depends on cortical representations. Second, developmental findings show that semantic generalization can precede the formation of detailed episodic memories \citep{keresztes2018hippocampal}. Third, studies of narrative perception reveal that event representations are constructed within hierarchical cortical structures \citep{baldassano2017discovering,chang2022information}. Fourth, from a computational standpoint, processing complex stimuli such as images or words requires sophisticated cortical processing, making it implausible that episodic encoding could bypass these processes. A related limitation of most existing models is their assumption of a strictly unidirectional relationship between episodic and semantic memory, where the former merely trains the latter. This view struggles to account for phenomena such as semantic intrusions and gist-based distortions of memories, which are ubiquitous in episodic memory tasks \citep{schacter2011memory}.  

Most critically, existing frameworks fail to jointly explain the core empirical phenomena that characterize both semantic and episodic memory. Semantic memory underlies the generalization of learned associations (e.g., 3–red, 5–blue, 3–blue) to novel combinations (e.g., 5–red), reflecting mechanisms of statistical learning \citep{ngo2021contingency}. It is also sensitive to the frequency and typicality of stimuli and categories \citep{mccloskey1980stimulus,rosch1976structural}. By contrast, episodic memory supports the "mental search" of past experiences. Recognition and free-recall studies—such as judging whether an image was previously seen \citep{brady2008visual} or recalling words from lists \citep{kahana2020computational}—reveal robust behavioral regularities, including recency and serial-order effects, as well as semantic intrusions during retrieval. Moreover, episodic memory enables the constructive recombination of past experiences to generate novel scenarios, as shown in research on constructive episodic simulation and episodic future thinking \citep{schacter2017episodic,schacter2015episodic}.

None of the existing models of semantic–episodic interaction can account for the full range of phenomena across both domains. Moreover, they typically lack key mechanisms—such as the separation of content and context, and cue-dependent retrieval—that are central to leading models of memory search \citep{kahana2020computational,norman2008computational,atkinson1968human,howard2002distributed,farrell2002endogenous,sederberg2008context,raaijmakers1981search,gillund1984retrieval,shiffrin1997model,nelson2013co,sederberg2011human}. As a result, we still lack an integrative computational framework capable of jointly explaining the main empirical findings on semantic and episodic memory and their dynamic interactions.



In this study, we address this challenge by introducing a novel framework, the Generative Episodic–Semantic Integration System (GENESIS) model, which offers a new perspective on how semantic (cortical) and episodic (hippocampal) systems interact while reproducing key empirical findings from both domains.

\section{Results}

In this section, we introduce the GENESIS model (Section~\ref{sec:model}) and evaluate its performance in semantic tasks of statistical learning under capacity limitations and generalization (Section~\ref{sec:semantic}), as well as in episodic tasks involving recognition memory (Section~\ref{sec:episodic-recognition}), serial recall (Section~\ref{sec:episodic-serial-recall}), reconstruction under limited capacity (Section~\ref{sec:distortion}), and replay and recombination of episodes (Section~\ref{sec:replay}).

\subsection{The Generative Episodic–Semantic Integration System (GENESIS) model}
\label{sec:model}

\begin{figure}[H]
    \centering
    \includegraphics[width=0.9\textwidth]{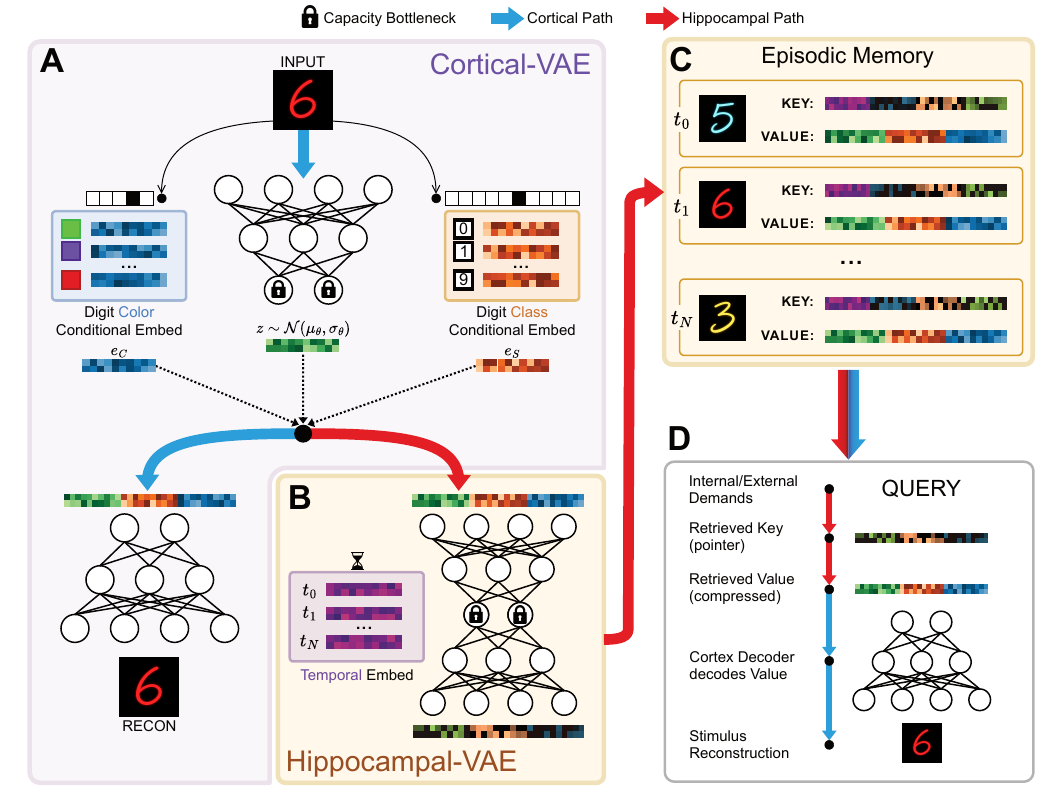}
    \caption{Illustration of the Generative Episodic–Semantic Integration System (GENESIS) model. The violet and yellow boxes denote core components of the semantic and episodic systems, respectively. Although shown separately, these systems closely interact, as illustrated in the white panel. \textbf{A.} The Cortical-VAE (a conditional $\beta$-VAE \citep{sohn2015learning}) achieves a representative input embedding by merging the Gaussian latent state $z$ with color and digit conditional embeddings, $e_C, e_S$, respectively, and provides the input to either the cortical decoder (blue arrow) or the Hippocampal-VAE (red arrow). \textbf{B.} The Hippocampal-VAE (a $\beta$-VAE \citep{kingma2013auto,higgins2017beta}) generates a compressed representation of the input embedding coming from the Cortical-VAE encoding, and incorporates temporal embeddings, to form an episode-specific key. \textbf{C.} Episodic memory is implemented as a Retrieval-Augmented Generation (RAG) system that memorizes episodes as key-value pairs (or sequences of key-value pairs). Values are item embeddings (possibly compressed by the Cortical-VAE). Keys are a mixture of item embeddings (further compressed by the Hippocampal-VAE) and temporal embeddings. \textbf{D.} Interaction between cortical and hippocampal systems given a query. The query triggers a query-key-value search process that retrieves one or more episodes from Episodic Memory and routes them to the decoder of the Cortical-VAE. See the main text for further explanation.}
    \label{fig.1}
\end{figure}

The Generative Episodic–Semantic Integration System (GENESIS) model is schematically illustrated in Figure~\ref{fig.1}. The model comprises two interconnected generative models (limited-capacity variational autoencoders or VAEs \citep{kingma2013auto,higgins2017beta}) -- a Cortical-VAE and a Hippocampal-VAE (Figure \ref{fig.1}) -- along with an Episodic Memory component, which we implement for simplicity using a Retrieval-Augmented Generation (RAG) system \citep{lewis2020retrieval}.

The model receives an input item (e.g., an image of the number 6 colored red, or 6-red) and first encodes it using the encoder of the Cortical-VAE, representing cortical processing. Because the encoder has limited capacity, it compresses the input into a latent item embedding comprising two class embeddings (for color and digit) and an item-specific latent vector $z$.  

This latent embedding is then processed along two parallel pathways (blue and red arrows in Figure \ref{fig.1}). First, it can be decoded by the Cortical-VAE’s decoder, corresponding to cortical (here, visual) reconstruction. Second, it can be routed to hippocampal circuits to form an episodic memory -- which boils down to creating a key–value pair. To do so, the item embedding is further encoded and decoded through the (limited-capacity) Hippocampal-VAE and combined with a temporal embedding that captures when the item was experienced. The resulting compressed representation forms the key, while the associated value corresponds to the Cortical-VAE item embedding. These key–value pairs are stored in a retrieval-augmented memory (RAG) architecture \citep{lewis2020retrieval}, where each pair constitutes an episode. More generally, an episode can also be viewed as a short sequence of such key–value pairs.  

Episodic recall operates via standard query–key matching within the RAG. A query vector is compared to stored keys via similarity metrics, and the most similar entries retrieve their corresponding values. Retrieved values are latent embeddings that can be decoded by the Cortical-VAE to reconstruct perceptual representations (e.g., images).  

As assumed by models of episodic search \citep{kahana2020computational}, queries initiating recall can arise from different sources in different psychological paradigms. In recognition (old/new) tasks, a probe image (e.g., a 3-yellow) is encoded by the Cortical-VAE; its embedding serves as the query. High query–key similarity indicates that the item was already known and low query–key similarity indicates that the item was new. In free or serial recall paradigms, recall is triggered by a temporal embedding corresponding to a specific time point, such as the moment when the participant is asked to recall. Existing models of episodic search sometimes extend the temporal embedding with a list identifier, but we do not use this mechanism in our simulations for simplicity. Retrieval can proceed iteratively, with the key of each recalled item serving as the query for the next, thereby generating a temporal sequence of remembered episodes.

In summary, although Figure~\ref{fig.1} depicts the core components of the episodic system (yellow boxes) as distinct from those of the semantic system (violet box), both episodic memory formation and recall arise from their interaction. Specifically, they depend on the Cortical-VAE (providing encoding and decoding functions), the Hippocampal-VAE (generating keys), and the RAG module (supporting query–key–value matching). Accordingly, semantic and episodic processes are not strictly separated or localized within isolated modules, but emerge from coordinated interactions among these components.


\subsection{Statistical learning, capacity limitations, and generalization in semantic memory}
\label{sec:semantic}

A central tenet of most accounts of semantic memory, including ours, is that it implements a form of statistical learning by gradually accumulating structured knowledge from experience—either directly perceived or replayed from episodic memory \citep{mcclelland1995there,schapiro2017complementary}. Here, we test the capacity of the Cortical-VAE to learn the statistical regularities of input images and to generalize by recombining learned elements in novel ways. Moreover, in line with evidence that semantic processing is limited by attention and memory \citep{marois2005capacity,zenon2019information}, we examine how encoding capacity in the Cortical-VAE shapes reconstruction quality and representational geometry.

To implement the Cortical-VAE, we employ a convolutional conditional $\beta$-VAE with limited capacity \citep{higgins2017beta,burgess2018understanding,sohn2015learning}. This deep generative model includes an encoder, which compresses an input (e.g., an image) into a latent representation, and a decoder, which reconstructs the input from that latent representation (Figure \ref{fig.1}A). The Cortical-VAE was trained in an unsupervised manner on colored MNIST digits to encode and reconstruct images. During encoding, each input image is mapped to a composite item embedding $[z, e_C, e_S]$, where $z$ is a Gaussian latent vector capturing item-specific variability, and $e_C$ and $e_S$ are discrete embeddings representing color and digit identity, respectively. 

\begin{figure}[H]
    \centering
    \includegraphics[width=0.95\textwidth]{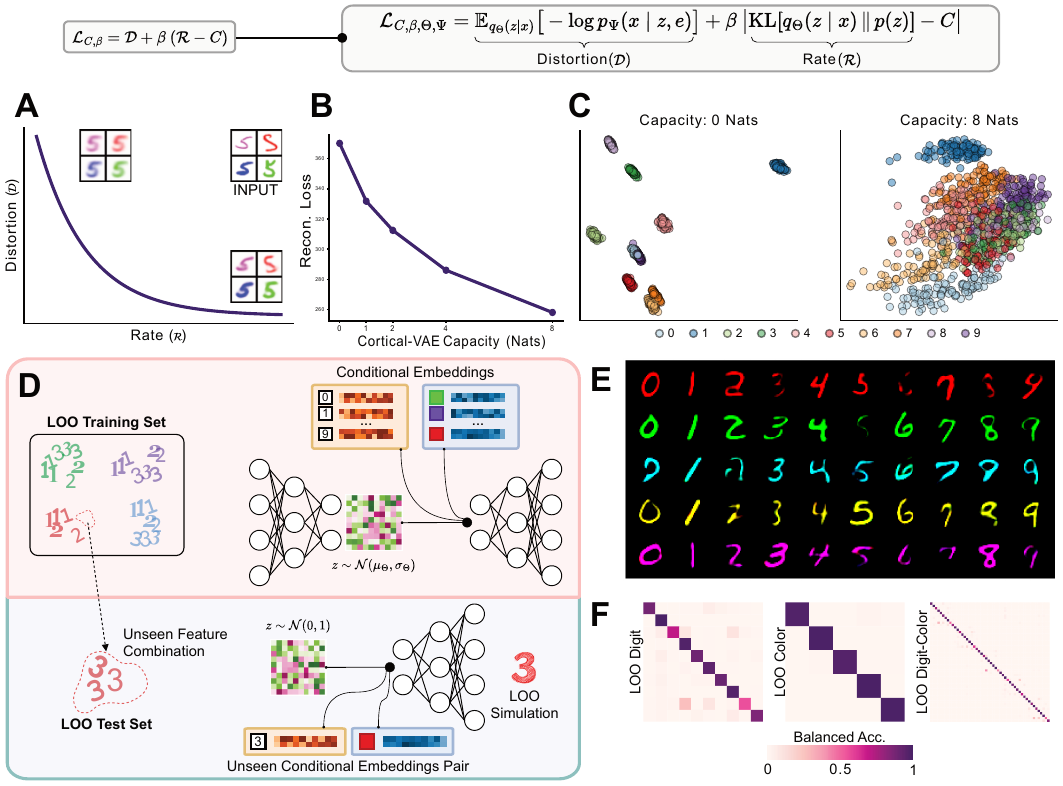}
    \caption{Simulation of statistical learning, capacity limitations, and generalization in semantic memory. \textbf{A.} Theoretical rate-distortion curve \citep{shannon1959coding}. The conditional $\beta$-VAE enables the implementation of the rate-distortion trade-off in the loss terms, as shown in the formula at the top ($e=[e_S, e_C]$ is the set of conditioning embeddings) \textbf{B.} Reconstruction performance of images as a function of Cortical-VAE capacity levels \textbf{C.} Multidimensional Scaling (MDS) of the latent feature maps of images (only red digits for simplicity) processed by the Cortical-VAE under low-capacity (0 nats, left) and high-capacity (8 nats, right). Feature maps are extracted from the first decoder layer. \textbf{D.} Schematic of the Leave-One-(Pair)-Out (LOO) protocol. The Cortical-VAE is trained on an ablated dataset with one digit–color pair (e.g., 5-red) withheld; the missing pair is later generated by combining the corresponding conditional embeddings. \textbf{E.} Simulated exemplars for each left-out digit–color pair. Each grid cell shows examples generated from a model trained with that specific pair ablated. \textbf{F.} Heatmaps indicating the classification accuracy of the generated digit–color pairs. See the main text for further explanation.}
    \label{fig.2}
\end{figure}

To account for limited capacity in the Cortical-VAE, we leverage the fact that the $\beta$ parameter of the $\beta$-VAE has a well-defined theoretical interpretation as a capacity bottleneck within rate–distortion theory \citep{shannon1959coding}, where decreases in reconstruction accuracy with limited resources follow a characteristic rate–distortion curve (Figure~\ref{fig.2}A). This framework treats encoding capacity as a proxy for available cognitive resources (e.g., attention, memory) during learning, which may vary with cognitive demands (e.g., performing a dual task may reduce capacity) or contextual factors (e.g., salient or surprising events may recruit greater capacity) \citep{nagy2025adaptive,d2025geometry,bates2020efficient,bhui2021resource,sims2016rate,sims2003implications,sims2018efficient,gershman2021rational,jakob2023rate,anderson1989human,pezzulo2024neural}.

To systematically examine how limited encoding capacity in the Cortical-VAE affects reconstruction quality, we trained the model under varying capacity constraints (${0, 1, 2, 4, 8}$ nats). As expected, the model trained at high capacity accurately reconstructed MNIST images. This finding aligns with extensive prior work showing that $\beta$-VAEs support robust statistical learning of visual stimuli, whether trained on real inputs or through generative replay \citep{higgins2017beta,burgess2018understanding,shin2017continual,van2020brain,stoianov2022hippocampal}. Reducing capacity led to a progressive decline in reconstruction accuracy, following the characteristic rate–distortion curve of rate–distortion theory (Figure~\ref{fig.2}B). Interestingly, high- and low-capacity training also produced distinct latent geometries. Figure~\ref{fig.2}B–C compares the first-layer decoder representations at high capacity (8 nats, right) and low capacity (0 nats, left). At high capacity, representations of different instances of the same digit (colored dots) spanned a wide region of latent space, whereas at low capacity they became less differentiated and collapsed into a compact region, with all instances converging toward a single prototype representation.

We next tested the high-capacity Cortical-VAE in a more challenging semantic generalization task, which involves generating novel, unseen digit-color pairs. For this, we implemented a Leave-One-(Pair)-Out (LOO) protocol, using the high-capacity Cortical-VAE (Figure~\ref{fig.2}D). In each of 50 experiments, all samples from one digit–color pair (e.g., all 5-red images) were withheld during training. During testing, the Cortical-VAE synthesized the missing category by combining the corresponding conditional embeddings (e.g., 5 and red) and sampling a random $z$. As shown in Figure~\ref{fig.2}E, the model successfully generated plausible images of previously unseen digit–color combinations. To evaluate these generated images, we trained independent classifiers (a CNN feature extractor with an MLP classification head on top to recognize digit, color, and digit–color pairs from the colored MNIST dataset. The classifiers accurately identified the unseen categories, confirming that the Cortical-VAE achieved strong compositional generalization in this domain (Figure~\ref{fig.2}F).


Summing up, these results demonstrate that the Cortical-VAE captures key properties of semantic memory, including statistical learning, the modulation of performance and representational geometry by encoding capacity in line with rate–distortion theory, and generalization to novel stimulus combinations.

\subsection{Recognition memory}
\label{sec:episodic-recognition}

A key function of the episodic system is to support recognition memory, which in its most common form involves determining whether an input (e.g., an image) has been previously experienced and is therefore familiar. This ``old/new'' paradigm has been extensively used in studies with words and images, showing that retrieval success is associated with the replay of episodic content \citep{wimmer2020episodic} and that performance declines with increasing list length \citep{kahana2020computational,howard2018memory}.  

\begin{figure}[H]
    \centering
    \includegraphics[width=0.9\textwidth]{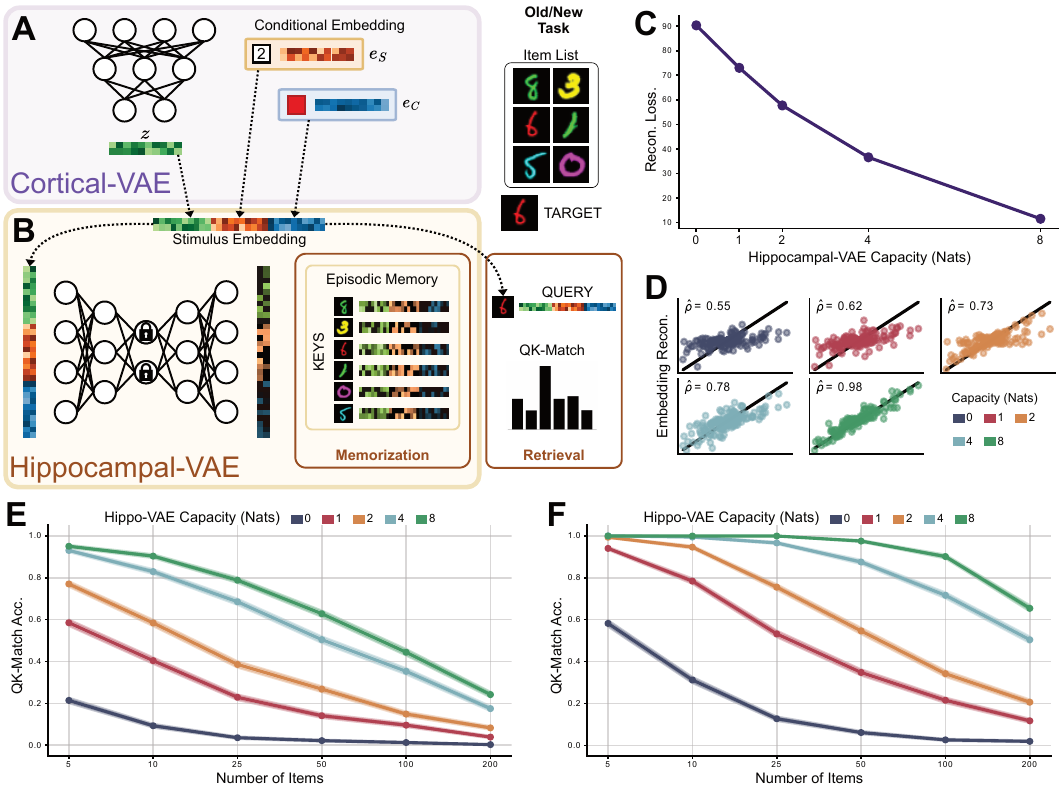}
    \caption{Simulations of recognition memory tasks. \textbf{A.} Encoding in the Cortical-VAE to obtain the item embedding $[z, e_S, e_C]$, where here $z=\mu_\Theta$. \textbf{B.} Memory search mechanism during the "old–new" task. Item embeddings from the target list are compressed by the Hippocampal-VAE and memorized in Episodic Memory. During retrieval, a query–key match is performed (temporal embeddings are omitted here, as they were not used), where the query corresponds to an item that may or may not have appeared in the list. \textbf{C,D.} Reconstruction performance of item embeddings as a function of Hippocampal-VAE capacity levels. \textbf{E.} Query–key match accuracy in the old–new task under the top-1 criterion, showing success rate as a function of list size and Hippocampal-VAE capacity. A success occurs when the query matches the closest key by cosine similarity. \textbf{F.} Query–key match accuracy under the top-3 criterion, where a success occurs if the correct key is retrieved within the three closest matches. See the main text for further explanation.}
    \label{fig.3}
\end{figure}

In the GENESIS model, the episodic system comprises two components: the Hippocampal-VAE and the Episodic Memory. To implement the Hippocampal-VAE, we employ a capacity-limited $\beta$-VAE \citep{kingma2013auto,higgins2017beta} trained to reconstruct the item embeddings generated by the Cortical-VAE. The output of the Hippocampal-VAE constitutes a key that captures a compressed episodic trace of the cortical representation (Figure~\ref{fig.3}A--B), along with a temporal embedding that we omit from this simulation for simplicity. Each key is stored together with its corresponding value (the item embedding generated by the Cortical-VAE) in the Episodic Memory.  

Given that the Hippocampal-VAE is a capacity-limited system, we first evaluated its reconstruction performance across five levels of representational capacity (${0, 1, 2, 4, 8}$ nats). As expected, reconstruction error increased smoothly with decreasing capacity, following a characteristic rate–distortion curve (Figure~\ref{fig.3}C–D).

Next, we implemented a recognition memory (``old/new'') experiment in which the GENESIS model was first presented with a list of items to memorize, followed by probe images to test whether they were correctly recognized as ``old,'' i.e., already present in Episodic Memory. Recognition was based on a query--key matching mechanism: the probe embedding served as the query, and its cosine similarity with all stored keys was computed. The model correctly recognized an item as ``old'' if the best-matching key corresponded to the target (top-1 accuracy; Figure~\ref{fig.3}E) or was among the three best matches (top-3 accuracy; Figure~\ref{fig.3}F). Two factors were manipulated: the capacity of the Hippocampal-VAE (${0, 1, 2, 4, 8}$ nats) and the list length ($5, 10, 25, 50, 100, 200$ items).  

Our results show that increasing list length decreases recognition accuracy, consistent with extensive empirical evidence \citep{kahana2020computational}. Moreover, a trade-off emerges between compression and discriminability: as the capacity of the Hippocampal-VAE decreases, both the intercept (baseline accuracy) and the slope (rate of accuracy decline with list size) decrease. This finding highlights that when keys are encoded with low capacity, the associated episodic memories become more difficult to retrieve—a difficulty that is amplified as the number of stored memories increases, reflecting reduced distinctiveness among key representations.

Summing up, these results show that the Hippocampal-VAE supports recognition memory in a way that is modulated by both cognitive resources and task demands (i.e., the number of items to be memorized). Notably, recognition in the model relies solely on the query--key matching mechanism and does not require reconstructing (decoding) the input. This indicates that the primary computational load of memory search operates on compressed keys, thereby conserving cognitive resources. It also implies a functional distinction between the relatively simple task of determining whether an item has been previously experienced (as requested by the "old/new" task) and the more demanding task of reconstructing its content.

\subsection{Serial recall}
\label{sec:episodic-serial-recall}

Another classic paradigm for studying episodic memory is serial recall. In this task, participants are presented with a list (or multiple lists) of items—typically images or words—and are later asked to reproduce them in the original order. This paradigm consistently reveal robust behavioral regularities, including serial order effects—items tend to be recalled in the order of presentation—as well as recency and, under some conditions, primacy effects \citep{kahana2020computational,murdock1962serial,postman1965short}.

\begin{figure}[H]
    \centering
    \includegraphics[width=0.8\textwidth]{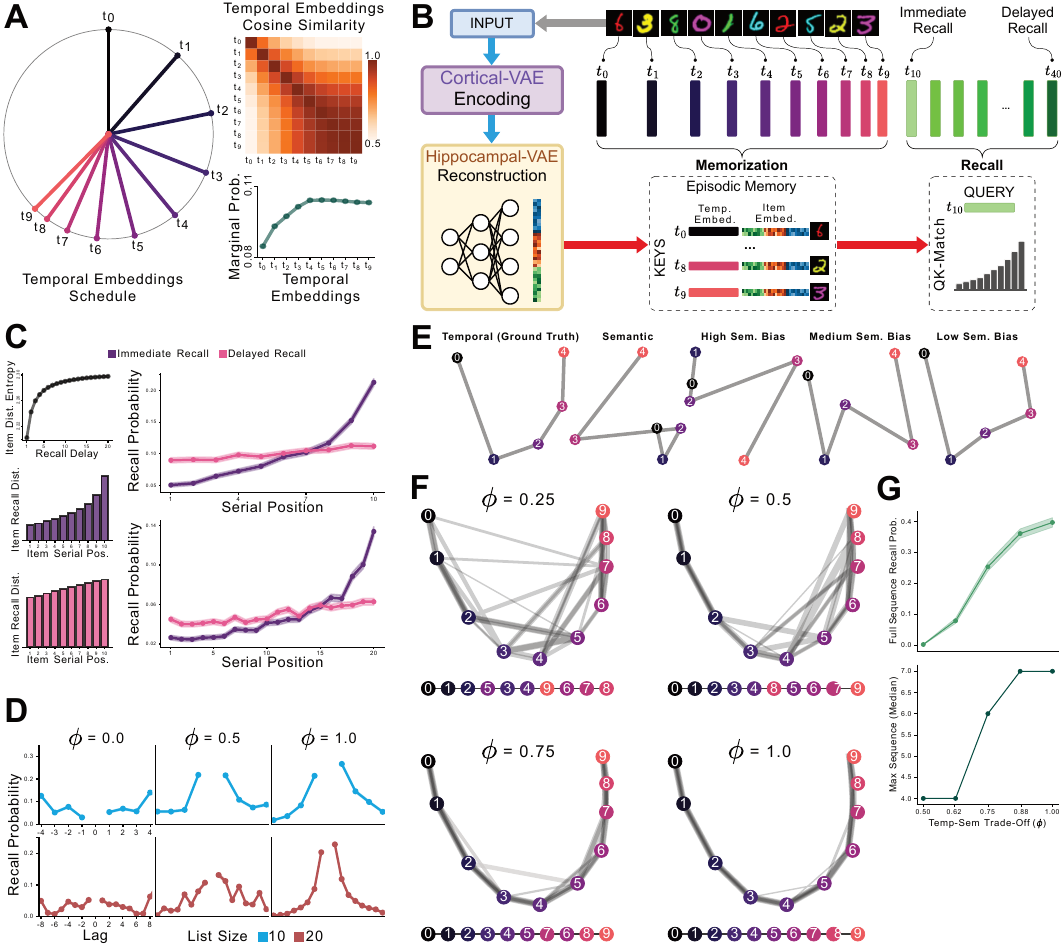}
    \caption{Simulations of serial recall tasks. \textbf{A.} A conceptual illustration of the temporal embeddings generation schedule, where the geometric distance between subsequent temporal embeddings decreases as time unfolds. \textbf{B.} Schema describing the usage of the temporal embeddings in the Episodic Memory, as items are learned through time. Internal time representations shape both item-related keys in Episodic Memory and queries at the time of recall. \textbf{C.} Modulation of the recency effect as a function of recall time. The entropy of the recall probability distribution of the list items increases as the recall time is delayed. \textbf{D.} Illustration of the forward effect describing the bias towards positive lags when trying to recall the most remembered item based on a cue. \textbf{E.} Geometric representation of temporal and semantic distances between items in the list. Nodes represent an item presented at a given position $t_0, t_1, \ldots, t_N$. The grey lines represent the path connecting the items in the correct temporal order. \textbf{F.}  Serial recall memory search path starting from the first item presented in the list as a cue (the item at $t_0$). Nodes represent the temporal distance between the items in the list. The grey lines present the most likely sequential recall path for a given level of $\phi$. The horizontal nodes list at the bottom is the most likely node sequence recalled. \textbf{G.} Visualizations of the probability of recalling the full sequence in the correct order (top), and the median of the maximum sub-sequence recalled (bottom), as a function of $\phi$. See the main text for further explanation.} 
    \label{fig.4}
\end{figure}

To model serial recall of image lists, we extend the keys of the episodic item embeddings with temporal embeddings that index the moment each episode was experienced, with distinct temporal embeddings assigned to each list element. Figure~\ref{fig.4}A schematically illustrates the temporal embedding scheme. For successive items in a list, the temporal embedding of element $i+1$ is separated from that of element $i$ by a constant angular displacement $\alpha$ that decreases gradually with time. As a result, the temporal embedding of an item $i$ is more similar (i.e., has a smaller cosine distance) to that of its immediate successor than to other items in the list (Figure~\ref{fig.4}B). This procedure yields a smooth trajectory of embeddings that gradually converge, providing a structured representation of temporal evolution and naturally giving rise to the forward bias widely observed in serial recall \citep{kahana1996associative}.

We next simulated immediate and delayed serial recall (Figure~\ref{fig.4}C) by using a temporal embedding as the retrieval cue and varying its proximity to the embedding of the final list item. In the immediate condition, the query was close to the most recent temporal embedding (1 timestep lag), whereas in the ,delayed condition it was farther apart (20 timesteps lag), simulating contextual drift over time. The model reproduced a clear recency effect in the immediate condition, with the final list items recalled with higher probability. In contrast, this effect was markedly attenuated in the delayed condition, reflecting reduced overlap between the retrieval cue and the stored temporal contexts. These patterns were observed for both 10- and 20-item lists. This result aligns with empirical evidence showing attenuation of recency effects under delayed recall conditions \citep{kahana2020computational}.

During serial recall, we also manipulated a parameter $\phi$ controlling the relative contribution of the temporal embedding to the key representation. When the temporal component was dominant ($\phi = 1$), the model exhibited a strong serial order effect (Figure~\ref{fig.4}D): following recall of item $i$, the next most likely retrieval was item $i+1$. As $\phi$ decreased ($\phi = 0.5$), the serial order effect weakened, disappearing entirely when the temporal embedding was omitted ($\phi = 0$). These effects were consistent across lists of 10 and 20 items. This finding parallels empirical and computational evidence that temporal context reinstatement underlies serial ordering in human recall \citep{kahana2020computational,howard2002distributed}.

We next simulated a cued serial recall task with 5 items, in which the first item of the list was used as a cue (Figure~\ref{fig.4}E-F), to investigate how modulating the contribution of temporal versus semantic information affects recall dynamics. In this task, sequential recall can be interpreted as traversing a path over a temporal manifold. Figure~\ref{fig.4}E illustrates the ground-truth temporal structure of the path, derived from the temporal distances between successive items, and compares it to the case in which the item (semantic) embeddings are considered, with various weights. Each node represents the temporal index of an item in the list (e.g., items 0, 1, 2, 3 and 4 are the first, second, third, fourth and last list items). The nodes are positioned according to either temporal or semantic similarity, and the paths connect the nodes sequentially in the order of presentation.

Figure~\ref{fig.4}F shows the results of a cued serial recall task with 10 items, overlaying the simulated recall paths obtained for different values of $\phi$, plotted on the geometric layout of nodes defined by the ground truth. The correct path is the one that follows the minimal temporal distances between nodes (0–1–2–...–9). Thicker paths correspond to trajectories sampled with higher probability, while darker paths indicate overlapping trajectories. As $\phi$ increases, meaning that the model becomes more driven by the temporal embedding, paths that connect close nodes (e.g., 1 and 2) are increasingly selected, reflecting correct serial recall. Conversely, as $\phi$ decreases, the model increasingly selects paths connecting more distant nodes (e.g., 1 and 7), reflecting semantic intrusions. This result aligns with evidence that semantically related stimuli (e.g., conceptually similar images or words) produce higher rates of recall errors in serial memory tasks \citep{kahana2020computational}.

Figure~\ref{fig.4}G summarizes model performance as a function of $\phi$, showing (left) the probability of successfully recalling the entire sequence, and (right) the longest correctly recalled subsequence. Both measures improve monotonically with increasing $\phi$, confirming that stronger temporal weighting facilitates ordered recall along the correct path.

Summing up, these results show that the Hippocampal-VAE supports serial recall and produces various widely reported findings, including the recency effect and its attenuation with distal recall, serial order effects, and semantic intrusions when the temporal embedding is down-weighted. 

\subsection{Gist-based distortion of episodic memory under low-capacity encoding}
\label{sec:distortion}

Prior semantic knowledge influences episodic processing in multiple ways. In the previous section, we illustrated the case of semantic intrusions, in which episodic retrieval produces items not present in the original list. A second way in which semantic memory affects episodic processing is by distorting the retrieved content. A well-known example is gist-based distortion, in which a retrieved memory becomes biased toward the semantic prototype of a category, reflecting an average or typical item rather than the specific episode that was encoded \citep{schacter2011memory,hemmer2009integrating,tompary2021semantic}.

\begin{figure}[H]
    \centering
    \includegraphics[width=0.8\textwidth]{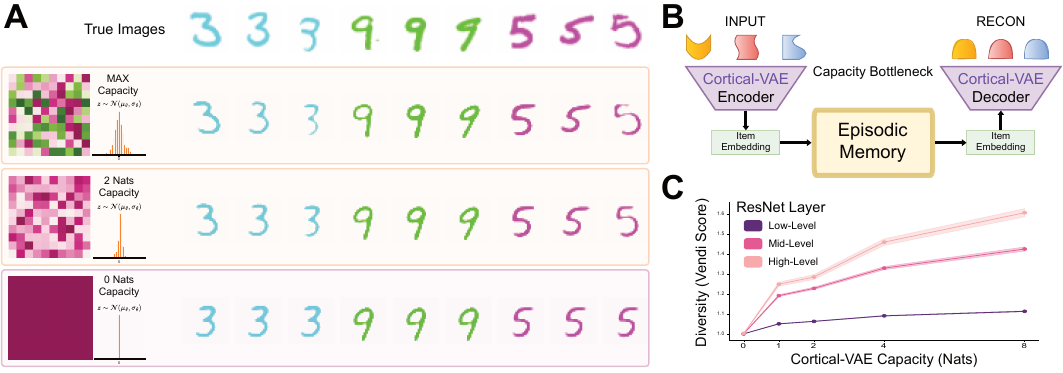}
\caption{Simulations of gist-based distortion of episodic memory under low-capacity encoding. \textbf{A.} Examples of stimuli reconstructed by the Cortical-VAE from episodic memories encoded with varying capacity levels, showing a progressive shift toward gist-like, prototypical representations as capacity decreases. \textbf{B.} Simulation procedure: image encoding at different capacity levels using the Cortical-VAE encoder, memorization in Episodic Memory, and decoding with the Cortical-VAE decoder. \textbf{C.} Quantification of reconstruction self-similarity using a diversity (\textit{Vendi}) score computed over ResNet feature maps extracted from images reconstructed at different capacity levels (see Section~\ref{sec:methods} for details). See the main text for further explanation.}    
    \label{fig.5}
\end{figure}

Here, we examined whether capacity limitations during episodic encoding can lead to such gist-based distortions. The rationale is that when episodes are encoded by the Cortical-VAE under low capacity, fine-grained details are lost, and reconstruction through the Cortical-VAE decoder may rely more heavily on schematic or semantic regularities captured during training. To test this hypothesis, we analyzed the model’s reconstruction behavior using item embeddings corresponding to episodic memories encoded by the Cortical-VAE under different capacity constraints (${0, 1, 2, 4, 8}$ nats) (Figure~\ref{fig.5}B). As expected, episodic memories encoded with high capacity were reconstructed nearly perfectly, preserving unique perceptual details. In contrast, when encoding capacity decreased, reconstructions of different exemplars within the same digit–color category became increasingly similar to one another (Figure~\ref{fig.5}A). We quantified this effect using the \textit{Vendi Score}, computed over ResNet embeddings \citep{he2016deep}, which decreased as reconstructed items became more self-similar (Figure~\ref{fig.5}C). This is equivalent to reconstructing an image from a latent state $z$ lying at the mean of the Gaussian prior ($z=\boldsymbol{0}$). Although the images are reconstructed based on episodic memories, they lack episodic details and correspond to prototypes that reflect prior semantic knowledge. 

Summing up, these results indicate that limited encoding capacity in the Cortical-VAE induces gist-based distortions during retrieval: reconstructed memories lose episodic specificity and converge toward semantic prototypes.

\subsection{Replay and recombination of episodes for constructive episodic simulation}
\label{sec:replay}

The hippocampus "replays" memories during offline periods, such as rest and sleep \citep{skaggs1996replay,foster2017replay,schuck2019sequential}. This memory reinstatement is thought to support various cognitive functions, such as planning and imagination \citep{pfeiffer2013hippocampal,pezzulo2017internally,olafsdottir2018role,redish2016vicarious,buckner2010role,lin2025neural}, episodic future thinking \citep{schacter2017episodic,schacter2015episodic}, and the training of the brain's generative models and policies \citep{mcclelland1995there,lorincz2000two,singh2022model,pezzulo2021secret,mattar2018prioritized,stoianov2022hippocampal,carr2011hippocampal}. 

\begin{figure}[H]
    \centering
    \includegraphics[width=0.8\textwidth]{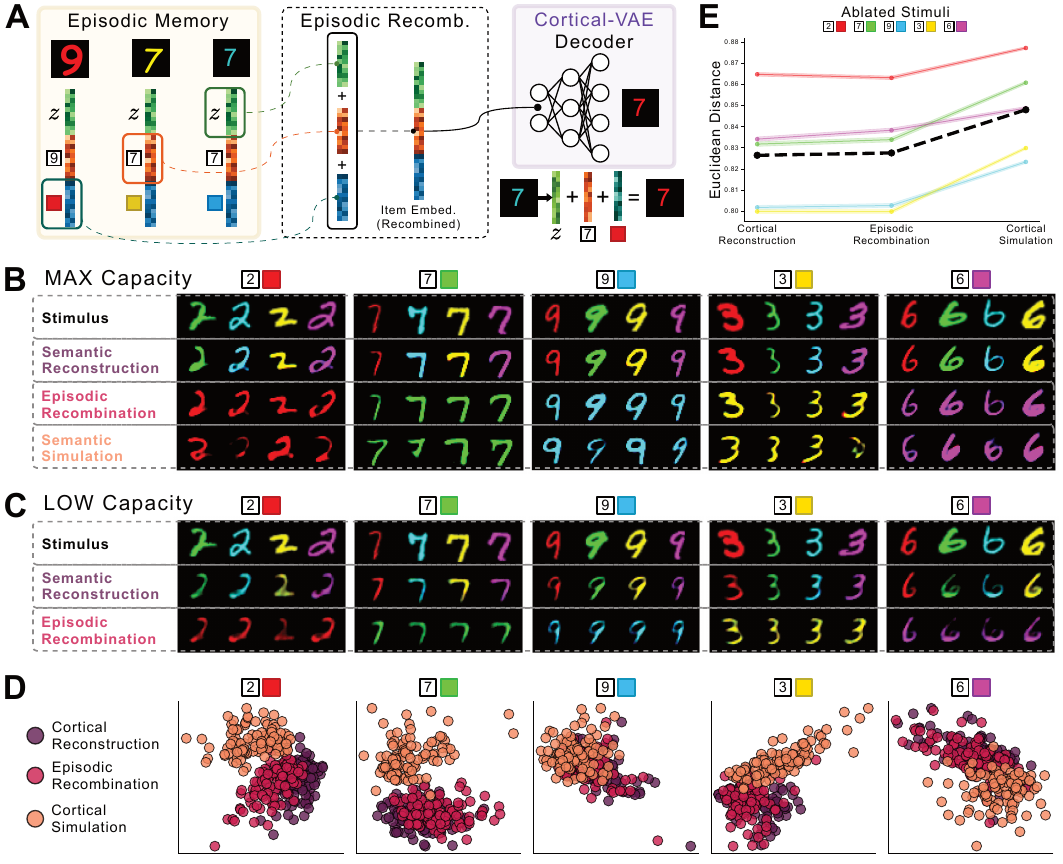}
    \caption{Simulations of episodic recombination and reconstruction.
    \textbf{A.} Episodic recombination procedure: a novel item embedding is constructed by combining the latent state $z=\mu_\Theta$, digit embedding $e_S$, and color embedding $e_C$ from three different stored items, and decoded by the Cortical-VAE. \textbf{B.} Comparison between semantic (Cortical-VAE) reconstruction of the input stimulus, episodic recombination, and semantic simulation of novel stimuli. Columns correspond to digit--color pairs ablated during training. Rows show (i) the original training stimulus, (ii) its semantic reconstruction, (iii) the reconstruction of an item embedding obtained via episodic recombination ($z$ and $e_S$ from the original stimulus, $e_C$ from the ablated embedding), and (iv) semantic simulation obtained by sampling $z$ while using the ablated $e_S$ and $e_C$. \textbf{C.} Same as in \textbf{B}, but using a low-capacity (1 nat) Cortical-VAE, highlighting the loss of episodic detail. \textbf{D.} Multidimensional scaling (MDS) of the generated images, computed on color-blind (Sobel-filtered) images where only edges/shapes are preserved, using high-level feature maps extracted from the third layer of ResNet. \textbf{E.} Quantification of reconstruction fidelity, also using Sobel-filtered representations. Average Euclidean distance between original stimuli and cortical-reconstructed, episodic-recombined, and cortical-simulated samples is reported relative to the max-capacity Cortical-VAE. See the main text for further explanation.}
    \label{fig.6}
\end{figure}

A particularly intriguing aspect of replay is that it is not limited to already experienced events, but can constructively recombine experiences in ways that have never been directly encountered. Indeed, there is substantial evidence in rodent and human studies that hippocampal replay sometimes includes novel, unexperienced events, such as "shortcuts" between experienced spatial trajectories \citep{gupta2010hippocampal} or sequences implied by learned abstract knowledge \citep{liu2019human}. The capability to recombine experiences has been linked to generative training \citep{stoianov2022hippocampal,van2020brain}, compositional inference in human problem solving \citep{schwartenbeck2023generative}, and dreaming \citep{deperrois2022learning,capone2024towards,hobson2012waking}. 

One way to test people’s ability to recombine episodic memories—also known as constructive episodic simulation \citep{schacter2007cognitive}—is to ask participants to recall autobiographical memories and then recombine elements from different memories to create never-experienced episodes \citep{addis2009constructive,addis2010episodic}.

Here, we examine whether GENESIS supports this form of episodic recombination, potentially linked to hippocampal replay. For this, we leverage the fact that the conditional $\beta$-VAE factorizes inputs into discrete feature components, such as digit identity ($e_S$) and color ($e_C$), which can be systematically recombined to form novel episodes. Specifically, we simulate recombination by selecting components from the item embeddings of known digit–color pairs from Episodic Memory (e.g., 9–red, 7–yellow, 7–blue), combining them to generate a novel embedding for an unseen pair (e.g., 7–red), and decoding it using the Cortical-VAE (Figure~\ref{fig.6}A,B).

To test the effectiveness of episodic recombination, we trained the Cortical-VAE by ablating five digit–color pairs from the training set (2–red, 7–green, 9–blue, 3–yellow, 6–magenta). Then, we performed three analyses comparing episodic recombination against two control conditions: semantic reconstruction (direct encoder–decoder reconstruction of test inputs) and semantic simulation (sampling new items from the Cortical-VAE using the ablated digit–color embeddings). 

In the first analysis, we asked whether episodic recombination yields new items that remain structurally faithful to the stimuli available in memory. For each ablated digit–color pair, we sampled 20 exemplars from related categories (e.g., 2–green, 2–blue, 2–yellow, 2–magenta) and combined their latent states $z$ with the embeddings of the ablated pair (e.g., 2-red). This procedure was repeated 500 times, producing a large set of recombined item embeddings that were passed to the cortical-VAE decoder to generate images. For each generated image, we computed the Euclidean distance matrix at the pixel level between the recombination-generated image and the original stimuli, averaging the distance scores across exemplars to obtain a population-level dissimilarity score. This procedure was repeated for the other two conditions, semantic reconstruction and semantic simulation, yielding a set of 500 Euclidean distance scores per ablated digit–color pair. A mixed-effects model with ablation pair as a random factor revealed that episodic recombinations were significantly closer to the originals than semantic simulations (coef $=0.033$, 95\% CI $[0.032, 0.034]$, $p<0.001$), as shown in Figure~\ref{fig.6}E. This confirms that recombination preserves structural details of episodic traces. This result can also be appreciated visually, by noticing that the shapes of the numbers in the novel pairs generated with episodic recombination—but not with semantic simulation—are very close to the original stimuli in Episodic Memory. This speaks to the fundamental difference between constructive episodic simulation, which maintains the unique characteristics of existing memories, and semantic simulation, which generates arbitrary new stimuli from the same distribution as the experienced stimuli \citep{schacter2007cognitive,suddendorf2007evolution,schacter2017episodic,schacter2015episodic}.

In the second analysis, we repeated the same procedure to compare episodic recombination, semantic reconstruction, and semantic simulation, but using a low-capacity ($1$ nat) Cortical-VAE encoder (Figure~\ref{fig.6}C). This simulation shows that episodic recombination based on memories encoded at low capacity yields novel stimuli that lose fine episodic details and become increasingly self-similar. These results demonstrate that capacity limitations affect not only the recall of past experiences but also their recombination during constructive episodic simulation. Speculatively, this focus on capacity limitations may provide a new perspective on why episodic recollection and recombination tend to lose detail in older adults \citep{addis2010episodic}.

In the third analysis, we examined the representational structure of the three conditions (episodic recombination, semantic reconstruction, and semantic simulation) at high capacity (Figure~\ref{fig.6}D). For each ablated pair, we generated a sample of 20 items per condition. Each item was encoded with a pre-trained ResNet \citep{he2016deep}, and feature maps extracted from the third layer were averaged to obtain centroids representing the group-level structure of each condition. This procedure was repeated 100 times, yielding a set of 100 group-level centroids per condition. We then applied multidimensional scaling (MDS) to the centroids to obtain a low-dimensional representation of their similarity relations. The resulting geometry, depicted in Figure~\ref{fig.6}D, showed a separation between conditions: cortical reconstructions clustered with episodic recombinations, whereas semantic simulations diverged the most. This demonstrates that recombination produces novel items that are more faithful to episodic structure than unconstrained cortical simulations. 

Summing up, these results show that the recombination of episodic memories—potentially linked to hippocampal replay—is a powerful mechanism for constructive episodic simulation and is modulated by capacity limitations. Speculatively, episodic recombination may also contribute to the formation of “false memories” when the recombined item embeddings are stored as new episodic traces.


\section{Discussion}

A cornerstone of cognitive science is the idea that the brain relies on two declarative memory systems—semantic and episodic—which interact in multiple ways \citep{tulving1972episodic,tulving2002episodic}. Classical models of semantic–episodic interaction, such as the influential complementary learning systems (CLS) theory \citep{mcclelland1995there,norman2003modeling}, focus on specific types of interaction, notably the rapid encoding of experiences in the hippocampus and their subsequent replay to gradually train cortical semantic representations. Despite the importance of this framework and its many recent extensions \citep{spens2024generative,spens2025hippocampo,fayyaz2022model,nagy2025adaptive}, we still lack a model that simultaneously accounts for key empirical findings in both semantic and episodic memory, as well as their dynamic interactions. 

In this study, we address this gap by introducing a novel framework—the Generative Episodic–Semantic Integration System (GENESIS)—which provides a unified account of how semantic (cortical) and episodic (hippocampal) systems interact, while simultaneously reproducing core empirical phenomena from both domains. At the core of the GENESIS model are two limited-capacity generative components, the Cortical-VAE and the Hippocampal-VAE, and an Episodic Memory component that stores and retrieves episodes through a query–key match mechanism. Together, these three components form an integrated system that enables multiple, bidirectional interactions between semantic and episodic subsystems.

Our simulations show that GENESIS provides an integrative account of findings from both semantic and episodic memory research, encompassing statistical learning and generalization (semantic), as well as recognition memory, serial recall, gist-based distortions and episodic recombination (episodic). Moreover, the model allows us to provide a normative perspective--grounded in rate-distortion theory--on how capacity limitations in the Cortical and Hippocampal VAEs shape the level of detail preserved in episodes and their memorability. The ability of GENESIS to capture these diverse phenomena within a single, generative architecture offers a promising step toward an integrative computational theory of declarative memory.

The GENESIS model introduces at least five elements of novelty compared to traditional frameworks addressing the relationship between semantic and episodic memory, such as the complementary learning systems (CLS) theory. First, episodic memories are not formed or recalled independently of semantic memory. Instead, the semantic system (Cortical-VAE) provides both the encoder (during memorization) and the decoder (during recall) for episodic representations. As shown in our simulations, this architecture allows semantic processing to shape episodic memory formation and retrieval—an interaction not captured by CLS or related models (but see \citep{fayyaz2022model,nagy2025adaptive,nagy2020optimal,spens2025hippocampo,jones2026}). This perspective is related to Tulving’s serial–parallel–independent (SPI) model \citep{tulving1995organization,tulving2001episodic,tulving1998episodic}, which proposes that perceptual information passes through the semantic system before being encoded in episodic memory. However, whereas the SPI model holds that the two systems operate independently during retrieval, in GENESIS episodic recall requires the semantic system.

Second, the contents of Episodic Memory are not the perceptual items themselves (e.g., images) but their latent embeddings, which must be decoded by the Cortical-VAE to reconstruct the original perceptual items. This design aligns with the indexing theory of hippocampal function, which posits that the hippocampus stores pointers to distributed cortical representations rather than full content \citep{teyler1986hippocampal,teyler2007hippocampal}, diverging from computational models that encode perceptual inputs in episodic components (e.g., the Hopfield network in \citep{spens2024generative}).

Third, episodic recall—or memory search—is implemented as a query–key–value matching process within a retrieval-augmented generator (RAG) framework \citep{lewis2020retrieval}. This mechanism preserves the separation of content and context central to leading models of episodic memory \citep{kahana2020computational,norman2008computational,atkinson1968human,howard2002distributed,farrell2002endogenous,sederberg2008context} and is central also in other recent accounts of memory search \citep{jimenez2024hipporag,gershman2025key}. A query may correspond to a temporal cue (as in free recall) or a semantic cue (as in spontaneous recollection). During sequential recall, the key of one retrieved element can serve as the query for the next, generating a chain of recalled episodes. Each key comprises two components: a semantic part, given by the compressed item embedding generated by the limited-capacity Hippocampal-VAE, and a temporal part, encoding when the episode occurred. Values correspond to the item embeddings produced by the Cortical-VAE encoder. Query–key similarity governs recall dynamics and naturally gives rise to semantic intrusions when items from different lists share high representational similarity. Importantly, memory search operates over the compressed keys rather than the full representations, offering an efficient and biologically plausible mechanism for retrieval under capacity constraints. Although query–key–value mechanisms are widely used in transformer-based attention models, where they typically require large amounts of training data, we adopt them here within a RAG framework. Once trained, the Hippocampal-VAE can generate episode-specific keys on the fly by compressing Cortical-VAE embeddings and integrating them with context-dependent temporal embeddings, a process compatible with rapid hippocampal learning.

Fourth, limited capacity—a ubiquitous property of biological generative systems—is a central feature of both the Cortical-VAE and the Hippocampal-VAE. In this respect, our model relates to theoretical work linking limited-capacity perception and memory to rate–distortion theory \citep{shannon1959coding} and adaptive information compression \citep{nagy2025adaptive,nagy2020optimal,spens2025hippocampo,d2025geometry,hemmer2009bayesian}. Crucially, GENESIS distinguishes between at least two types of information compression in memory. Capacity constraints in the Cortical-VAE compress item embeddings, yielding episodic memories with fewer details—resembling prototypes when decoded—and, as a side effect, making them less memorable. By contrast, capacity limits in the Hippocampal-VAE compress keys, producing episodic memories that are harder to retrieve but not necessarily less detailed (unless the values are themselves compressed or lose details with time). This dissociation has important implications for interpreting empirical findings and generates novel predictions (see below). 


Fifth, the model offers a novel computational mechanism for constructive episodic simulation \citep{schacter2007cognitive} not considered in previous models. This provides a novel perspective on how hippocampal replay may enable the generation of novel, unseen combinations of previous experiences. Taken together, these novel elements enable GENESIS to account for a broad range of empirical phenomena concerning semantic and episodic systems and their interactions, as illustrated in our simulations. Nonetheless, a systematic comparison between the assumptions and explanatory scope of GENESIS and other influential frameworks, including complementary learning systems, remains an important goal for future work.



The GENESIS model makes several novel predictions at the cognitive level that can be tested empirically. First, because episodic memories are encoded through the Cortical-VAE, the quality of recall should depend systematically on the representational capacity of semantic regions during encoding. Neural or behavioral markers of reduced cortical capacity (e.g., divided attention, low salience, strong expectation) should yield less detailed but more prototypical recollections—mirroring the compression–fidelity trade-off described by rate–distortion theory \citep{shannon1959coding,d2025geometry}. Second, the model predicts that the hippocampus forms “indexes,” in line with indexing theory \citep{teyler1986hippocampal}, by compressing latent cortical embeddings through the Hippocampal-VAE and integrating them with additional embeddings, such as temporal representations and potentially others (e.g., spatial context embeddings). Future studies could examine whether the hippocampus implements this form of compression or whether indexes are constructed through alternative mechanisms. Third, the model predicts distinct memory failures depending on whether compression occurs in the Cortical-VAE or Hippocampal-VAE. Reduced cortical capacity should produce gist-based distortions, reflecting stronger semantic influence, whereas reduced hippocampal capacity should impair retrieval success without altering the perceptual quality of reconstructions. These effects could be tested by manipulating representational load or inhibition separately in cortical and hippocampal regions. Another prediction is that certain tasks, such as the “old/new” recognition task, require synergistic contributions of cortical processes (to encode the item) and hippocampal processes (to perform the query–key match). Crucially, however, successful performance does not necessarily require retrieval of the associated value. This prediction could be tested by designing experiments that dissociate the cognitive and neural signatures of query–key matching from those of full value retrieval. A further prediction is that in certain cases, the task could be shortcut by the Cortical-VAE computing a coarse-grained sense of "familiarity" based on the similarity of the item to its learned semantic representations, without full episodic retrieval---but this sense of familiarity would lack episodic details. Finally, the model predicts that replay and generative recombination can give rise to novel episodic constructions—such as imagined future events—whose plausibility and novelty are shaped by the statistical structure of the learned semantic space. Accordingly, neural or behavioral measures of creative recombination (e.g., in episodic future thinking) should reveal systematic effects of semantic similarity and compression on the balance between familiarity and originality in generated episodes.




This study has several limitations that open promising avenues for future research. First, for simplicity, the Cortical-VAE was pre-trained rather than co-trained with the episodic system. However, this account does not capture the richer interplay between memory systems during consolidation, in which the hippocampus guides the reorganization of semantic information in the cortex \citep{squire1992memory,squire2015memory,nadel1997memory,moscovitch2005functional}. The complementary learning systems framework proposes that the fast-learning hippocampal system replays experiences to train the slower-learning cortical system, leading to the gradual semanticization of episodic traces \citep{mcclelland1995there}. Relatedly, the transformation hypothesis suggests that hippocampally dependent episodic memories evolve into semantic or gist-like representations supported by cortical structures; although the hippocampus becomes unnecessary for retrieving semantic memories, it remains critical for episodic recollection \citep{winocur2011memory}. While GENESIS does not currently implement these mechanisms, they could be incorporated by using episodic retrieval and replay from the Hippocampal-VAE, as illustrated in Section \ref{sec:episodic-serial-recall}, to train the Cortical-VAE via generative replay \citep{stoianov2022hippocampal,shin2017continual}. Such an extension would establish a bidirectional interaction between episodic and semantic systems, accounting for both semanticization and the continued reliance on the Hippocampal-VAE for episodic retrieval even after semantic knowledge has formed.



Second, for simplicity, we have so far assumed that Episodic Memory stores episodes (values) as uncompressed item embeddings generated during encoding by the Cortical-VAE. However, one can also posit that, like keys, values themselves undergo lossy compression by the capacity-limited Hippocampal-VAE. Decoding such compressed values with the Cortical-VAE yields episodes that lose episodic detail and become increasingly chimeric at low capacity (see Figure~\ref{fig:hallucinations} for examples). Systematically examining these distortions and their potential links to phenomena such as dreaming and hallucination remains an important avenue for future research. 

Third, episodic memory was modeled as a hybrid system combining a generative Hippocampal-VAE with a passive retrieval-augmented memory (RAG). While this ensured conceptual clarity, real episodic memories are dynamic—subject to consolidation, reconsolidation, and decay \citep{hupbach2007reconsolidation,roediger2011critical,gershman2013neural,zaki2025offline}. Future extensions could make the RAG component adaptive through continual replay and re-encoding, emulating updating, consolidation, and forgetting. The decay of episodic memories over time could produce gist-based distortions during reconstruction, over and above those induced by limited-capacity encoding discussed in this study \citep{schacter2011memory}.

Fourth, future implementations could adopt more biologically realistic methods to form episodic memories, such as those based on behavioral time scale synaptic plasticity \citep{li2023rapid,wu2025simple} or modern Hopfield networks \citep{ramsauer2020hopfield,spens2024generative} and study how they organize experiences into cognitive maps \citep{behrens2018cognitive,george2021clone,bottini2020knowledge,schapiro2017complementary,van2024hierarchical}. Relatedly, our current implementation relies on machine learning tools such as VAEs, RAG, and query–key–value mechanisms. These capture key computational principles we consider central—compression, flexible retrieval, and episodic indexing—but only in an abstract form that departs from biological implementation. For example, key–value memory architectures, which distinguish stored content (values) from retrieval cues (keys), are increasingly recognized as important, yet future work should explore biologically grounded realizations of these mechanisms \citep{kozachkov2023building,tyulmankov2021biological,gershman2025key}.

Fifth, our simulations followed classical paradigms in which the model is externally instructed what to memorize (e.g., all items in a list). In natural settings, however, organisms autonomously decide when and what to store based on factors such as event segmentation, surprise, and saliency \citep{zacks2020event,kumar2023bayesian}. Future work could endow GENESIS with mechanisms for autonomous event boundary detection and memory updating \citep{lu2022neural}. Relatedly, future extensions could enable GENESIS to store not only episodes but also their associated utility, potentially allowing the model to capture episodic influences on decision-making \citep{nicholas2025proactive,nicholas2026episodic,zhou2024episodic,bornstein2017reminders,mason2025explicit}.

Sixth, recall dynamics were simplified. In the serial recall task, we did not implement a stopping rule, although recall empirically tends to terminate when confidence declines \citep{atkinson1968human,kahana2020computational}. Future implementations could include a confidence-based termination mechanism, where recall ceases when query–key similarity drops below threshold. 

Seventh, we did not address primacy effects, which may arise from stronger encoding or context reinstatement for early items \citep{howard2002distributed}. GENESIS could model these by allowing encoding capacity to vary with attention and novelty.

Eighth, GENESIS currently abstracts away from the temporal structure of replay. Empirical data show that episodic retrieval involves temporally compressed reactivation \citep{wimmer2020episodic,howard2018memory}, with latency scaling with temporal distance \citep{hacker1980speed,muter1979response}. Future studies could map cosine dissimilarities between temporal embeddings onto retrieval latency to model temporal dynamics of mental search and mental time travel \citep{suddendorf2007evolution,hills2015exploration}. It would also be possible to model sequential replay in GENESIS as a cascade of episodic retrievals, by chaining multiple query–key–value matches to generate context-dependent, temporally ordered reactivations of past events \citep{zhou2026unifying,stoianov2022hippocampal}.

Ninth, future developments of GENESIS should pursue a more detailed mapping between its computational components and the neurobiological mechanisms underlying semantic and episodic memory. For simplicity, we labeled the components of GENESIS according to the classical distinction between a fast-learning hippocampal episodic system and a slow-learning cortical semantic system. However, this dichotomy is overly simplistic. Converging evidence indicates that semantic memory also depends on medial temporal lobe regions \citep{binder2011neurobiology,renoult2019knowing}, while episodic memory critically engages neocortical regions \citep{addis2018episodic,rugg2013brain,schacter2007remembering}. Accordingly, the present labels should be understood as heuristic rather than anatomical commitments. A key direction for future work is to clarify how the proposed components of GENESIS map onto interactive hippocampal–cortical dynamics supporting semantic and episodic memory.

Relatedly, semantic and episodic memory formation likely involve richer dynamics than currently captured by GENESIS. Contrary to the traditional view of a fast-learning hippocampal system and a slower-learning cortical system, evidence suggests that memory engrams can form rapidly in posterior parietal cortex, in a manner that appears partly hippocampus-independent \citep{brodt2016rapid,brodt2018fast}. Moreover, recent theoretical accounts propose that multiple representations of an episodic memory may be encoded simultaneously at different levels of granularity—from precise, context-specific details to more generalized, gist-like representations—and that semantic knowledge about specific events may form concurrently with the corresponding episodic memory rather than only afterward \citep{gilboa2021no,tarder2026adaptive}. These complexities fall beyond the current scope of GENESIS but represent important directions for future research.

Finally, future versions of GENESIS could test whether deficits in semantic (e.g., semantic dementia) or episodic (e.g., Alzheimer’s disease) processing can be explained in terms of capacity limitations \citep{patterson2007you}.


Summing up, the GENESIS model provides a unified computational framework that jointly accounts for core empirical phenomena in semantic and episodic memory while highlighting how generative mechanisms and capacity limitations shape memory encoding, retrieval, and reconstruction. By integrating two limited-capacity generative systems—the Cortical-VAE and the Hippocampal-VAE—within a retrieval-augmented architecture, the model offers a principled account of how experiences are encoded, recalled, and generalized across contexts. Beyond reproducing classical findings, GENESIS opens the way to a new generation of models that treat memory as an active, constructive, and resource-bounded process. Ultimately, GENESIS advances a view of declarative memory as a generative and constructive process, integrating semantic and episodic components within a unified computational framework.


\section{Methods}
\label{sec:methods}

\subsection{Cortical-VAE architecture and training}
The Cortical-VAE was implemented as a conditional $\beta$-VAE \citep{higgins2017beta,burgess2018understanding,sohn2015learning}, trained to reconstruct colored MNIST digits ($x \in \mathbb{R}^{3 \times H \times W}$). The dataset contained 50,000 training and 10,000 test samples, with each digit class evenly distributed across five colors. Images were resized to $H=W=32$. The encoder consisted of $m=3$ convolutional downsampling blocks, each with stride 2, kernel size 3, a residual block, and batch normalization, progressively reducing the input to spatial dimensions $h=H/2^m=4,\ w=W/2^m=4$. The decoder mirrored this structure with symmetric upsampling blocks. The latent representation was a Gaussian feature map $z \in \mathbb{R}^{16 \times 4 \times 4}$. Conditioning was implemented through two learnable embeddings, a digit embedding $e_S \in \mathbb{R}^{64}$ and a color embedding $e_C \in \mathbb{R}^{64}$, which were injected into the decoder via FiLM layers \citep{perez2018film} together with the latent feature map $z$. The complete architecture had approximately 561k parameters. Training was performed with AdamW (learning rate 0.0025), batch size 64, and 10 epochs. Binary cross-entropy was used as the reconstruction loss. To impose a limited-capacity bottleneck, we applied the $\beta$-VAE objective with $\beta=100$ and a warm-up schedule of 3,000 iterations to gradually reach the target capacity $C$. The loss function was
\[
\mathcal{L}_{C, \beta, \Theta, \Psi} = \mathbb{E}_{q_\Theta(z\mid x)} \big[-\log p_\Psi(x\mid z,e) \big] + \beta \, \big| \mathrm{KL}[q_\Theta(z\mid x)\,\|\,p(z)] - C \big|,
\]
where $\Theta$ and $\Psi$ denote encoder and decoder parameters, respectively, and $e=[e_S,e_C]$. To examine different bottleneck strengths, models were trained with target capacities $C \in \{0,1,2,4,8\}$ nats. See Figure \ref{fig:losses} of the Supplementary Materials for details.

\subsection{Hippocampal-VAE architecture and training}
The Hippocampal-VAE was implemented as a $\beta$-VAE, trained to reconstruct item embeddings produced by the Cortical-VAE. Each item embedding was defined as $h=[z,e_S,e_C]$, where $z \in \mathbb{R}^{16 \times 4 \times 4}$ was flattened into a 256-dimensional vector and concatenated with the digit and color embeddings ($e_S,e_C \in \mathbb{R}^{64}$), yielding $h \in \mathbb{R}^{384}$. Training data consisted of item embeddings derived from the same 50,000 training and 10,000 test images used for the Cortical-VAE. The encoder was a multilayer perceptron with three downsampling blocks: hidden layers of size 512 and 256 followed by a Gaussian latent layer of size 64. The decoder mirrored this structure with upsampling blocks of size 256 and 512. The complete architecture had approximately 593k parameters. Training was performed with AdamW (learning rate 0.001), batch size 32, and 8 epochs. Mean-squared error was used as the reconstruction loss. As in the Cortical-VAE, a limited-capacity bottleneck was enforced by applying the $\beta$-VAE objective with $\beta=100$ and a 3,000-iteration warm-up schedule to gradually reach the target capacity $C$. Models were trained with target capacities $C \in \{0,1,2,4,8\}$ nats.

\subsection{Leave-One-(Pair)-Out (LOO) protocol}

To test the generative flexibility of the Cortical-VAE, we implemented a Leave-One-(Pair)-Out (LOO) protocol. For simplicity, we used the Cortical-VAE trained unconstrained (by removing the target capacity from the loss). In each LOO run, the training set was modified to exclude a single digit–color pair. During generation, we sampled latent vectors $z \sim \mathcal{N}(0, I)$ and conditioned the decoder on the embeddings $e_S$ and $e_C$ corresponding to the withheld digit–color pair, i.e., a combination never observed together during training. This procedure was repeated exhaustively for all 50 digit–color pairs.

To evaluate whether the Cortical-VAE successfully extrapolated the missing combinations, we assessed the generated images using an external CNN classifier, trained separately to perform three classification tasks: (i) digit recognition, (ii) color recognition, and (iii) joint digit–color recognition. For each withheld digit–color pair, 100 generated images are fed to the classifier to compute logits.

\subsection{Progressive generation of temporal embeddings} 

Temporal embeddings $\{t_\tau\}$ were generated to ensure pairwise distances between the current and the next temporal embedding decrease as time $\tau$ unfolds. Therefore, each embedding was obtained by rotating the previous embedding by a scheduled angle. Given a temporal embedding $t_\tau$, the next embedding $t_{\tau+1}$ is constructed as  

\begin{equation}
t_{\tau+1} = \cos(\theta_\tau) \, t_\tau + \sin(\theta_\tau) \, t_\tau^{\perp},
\end{equation}  

where $t_\tau^{\perp}$ is a randomly sampled unit vector orthogonal to $t_\tau$, and $\theta_\tau$ is the scheduled angle controlling the similarity between successive embeddings.  

The series of angles $\{\theta_\tau\}$ was chosen such that the distance between consecutive embeddings decreases progressively over time. The sequence was generated using the Graham--Smith algorithm, which ensures controlled sampling of orientations on the unit hypersphere. A linearly decreasing $\theta_\tau$ schedule was adopted for all the experiments involving the temporal dimension. 

A temporal embedding $t_{\tau}$, together with an item embedding $h_{\tau}$ corresponding to a stimulus encoded at time $\tau$, define the key $k=[t_{\tau}, h_{\tau}]$ used in the Episodic Memory for serial and free recall tasks. The relative contribution of temporal versus item embeddings during memory search is controlled by the parameter $\phi$, such that:

$$ k= \left[ \phi \frac{t_{\tau}}{||t_{\tau}||}, \ (1-\phi) \frac{h_{\tau}}{||h_{\tau}||} \right]. $$

\subsection{Quantifying diversity in gist-based distorted image samples}

To examine whether low-capacity encoding introduces distortions in image reconstruction, we evaluated reconstruction performance using item embeddings from the Hippocampal-VAE corresponding to images originally encoded by Cortical-VAEs trained with $C \in {0, 1, 2, 4, 8}$. For each capacity level, we reconstructed the test images and analyzed whether degraded Gaussian latent states $z$ resulted in a loss of perceptual details. Specifically, we sampled 50 exemplars per digit–color combination from the test set, and quantified within-group diversity (or self-similarity). Lower scores indicate less image diversity, or increased self-similarity. Here, we assumed that a group of images with degraded perceptual details would result in high self-similarity.

Diversity was assessed using the Vendi score (VS; \citep{friedman2022vendi}), which measures the self-similarity of a set of embeddings $v=[v_1, v_2, \ldots, v_N]$:

$$\textrm{VS}(v_1, v_2, \ldots, v_N) = \exp \left(-\textstyle \sum_i^N\lambda_i \log \lambda_i \right)$$

where $\lambda_1, \ldots, \lambda_N$ are the eigenvalues of the normalized positive semidefinite similarity matrix $K/N$, with $K \in \mathbb{R}^{N \times N}$. To capture both low- and high-level perceptual diversity, we extracted embeddings from three layers of a pre-trained ResNet: layer 1 (low-level), layer 2 (mid-level), and layer 3 (high-level). For each layer, we computed $K$ as the cosine similarity matrix of the 50 feature maps extracted from the image group. The final Vendi score was obtained as the average of the scores computed across digit-color combinations, for each capacity level.

\section*{Acknowledgments}

This research received funding from the European Research Council under the Grant Agreement No. 820213 (ThinkAhead), the Italian National Recovery and Resilience Plan (NRRP), M4C2, funded by the European Union, NextGenerationEU (Project IR0000011, CUP B51E22000150006, “EBRAINS-Italy”; Project PE0000013, “FAIR”), and the Ministry of University and Research, PRIN PNRR P20224FESY and PRIN 20229Z7M8N. The funders had no role in study design, data collection and analysis, decision to publish, or preparation of the manuscript. We used a Generative AI model to correct typographical errors and edit language for clarity. 

\newpage

\bibliographystyle{apalike}  
\bibliography{references}  

@article{muter1979response,
  title={Response latencies in discriminations of recency.},
  author={Muter, Paul},
  journal={Journal of Experimental Psychology: Human Learning and Memory},
  volume={5},
  number={2},
  pages={160},
  year={1979},
  publisher={American Psychological Association}
}

@article{nagy2025adaptive,
  title={Adaptive compression as a unifying framework for episodic and semantic memory},
  author={Nagy, David G and Orb{\'a}n, Gerg{\H{o}} and Wu, Charley M},
  journal={Nature Reviews Psychology},
  pages={1--15},
  year={2025},
  publisher={Nature Publishing Group US New York}
}

@article{gupta2010hippocampal,
  title={Hippocampal replay is not a simple function of experience},
  author={Gupta, Anoopum S and Van Der Meer, Matthijs AA and Touretzky, David S and Redish, A David},
  journal={Neuron},
  volume={65},
  number={5},
  pages={695--705},
  year={2010},
  publisher={Elsevier}
}

@article{van2020brain,
  title={Brain-inspired replay for continual learning with artificial neural networks},
  author={Van de Ven, Gido M and Siegelmann, Hava T and Tolias, Andreas S},
  journal={Nature communications},
  volume={11},
  number={1},
  pages={4069},
  year={2020},
  publisher={Nature Publishing Group UK London}
}

@article{bates2020efficient,
  title={Efficient data compression in perception and perceptual memory.},
  author={Bates, Christopher J and Jacobs, Robert A},
  journal={Psychological review},
  volume={127},
  number={5},
  pages={891},
  year={2020},
  publisher={American Psychological Association}
}

@article{shannon1959coding,
  title={Coding theorems for a discrete source with a fidelity criterion},
  author={Shannon, Claude E and others},
  journal={IRE Nat. Conv. Rec},
  volume={4},
  number={142-163},
  pages={1},
  year={1959}
}

@article{d2025geometry,
  title={The geometry of efficient codes: How rate-distortion trade-offs distort the latent representations of generative models},
  author={D’Amato, Leo and Luca Lancia, Gian and Pezzulo, Giovanni},
  journal={PLOS Computational Biology},
  volume={21},
  number={5},
  pages={e1012952},
  year={2025},
  publisher={Public Library of Science San Francisco, CA USA}
}

@article{singh2022model,
  title={A model of autonomous interactions between hippocampus and neocortex driving sleep-dependent memory consolidation},
  author={Singh, Dhairyya and Norman, Kenneth A and Schapiro, Anna C},
  journal={Proceedings of the national academy of sciences},
  volume={119},
  number={44},
  pages={e2123432119},
  year={2022},
  publisher={National Academy of Sciences}
}

@inproceedings{higgins2017beta,
  title={beta-vae: Learning basic visual concepts with a constrained variational framework},
  author={Higgins, Irina and Matthey, Loic and Pal, Arka and Burgess, Christopher and Glorot, Xavier and Botvinick, Matthew and Mohamed, Shakir and Lerchner, Alexander},
  booktitle={International conference on learning representations},
  year={2017}
}

@article{shin2017continual,
  title={Continual learning with deep generative replay},
  author={Shin, Hanul and Lee, Jung Kwon and Kim, Jaehong and Kim, Jiwon},
  journal={Advances in neural information processing systems},
  volume={30},
  year={2017}
}

@article{spens2025hippocampo,
  title={Hippocampo-neocortical interaction as compressive retrieval-augmented generation},
author={Spens, Eleanor and Burgess, Neil},
journal={BiorXiv},
    year={2025}
}

@article{tyulmankov2021biological,
  title={Biological learning in key-value memory networks},
  author={Tyulmankov, Danil and Fang, Ching and Vadaparty, Annapurna and Yang, Guangyu Robert},
  journal={Advances in Neural Information Processing Systems},
  volume={34},
  pages={22247--22258},
  year={2021}
}

@article{kozachkov2023building,
  title={Building transformers from neurons and astrocytes},
  author={Kozachkov, Leo and Kastanenka, Ksenia V and Krotov, Dmitry},
  journal={Proceedings of the National Academy of Sciences},
  volume={120},
  number={34},
  pages={e2219150120},
  year={2023},
  publisher={National Academy of Sciences}
}

@article{zhou2026unifying,
  title={A unifying account of replay as context-driven memory reactivation},
  author={Zhou, Zhenglong and Kahana, Michael J and Schapiro, Anna C},
  journal={eLife},
  volume={13},
  pages={RP99931},
  year={2026},
  publisher={eLife Sciences Publications, Ltd}
}

@article{schwartenbeck2023generative,
  title={Generative replay underlies compositional inference in the hippocampal-prefrontal circuit},
  author={Schwartenbeck, Philipp and Baram, Alon and Liu, Yunzhe and Mark, Shirley and Muller, Timothy and Dolan, Raymond and Botvinick, Matthew and Kurth-Nelson, Zeb and Behrens, Timothy},
  journal={Cell},
  volume={186},
  number={22},
  pages={4885--4897},
  year={2023},
  publisher={Elsevier}
}

@article{lorincz2000two,
  title={Two-phase computational model training long-term memories in the entorhinal-hippocampal region},
  author={L{\"o}rincz, Andr{\'a}s and Buzs{\'a}ki, Gy{\"o}rgy},
  journal={Annals of the New York Academy of Sciences},
  volume={911},
  number={1},
  pages={83--111},
  year={2000},
  publisher={Wiley Online Library}
}

@article{foster2017replay,
  title={Replay comes of age},
  author={Foster, David J},
  journal={Annual review of neuroscience},
  volume={40},
  number={1},
  pages={581--602},
  year={2017},
  publisher={Annual Reviews}
}

@article{capone2024towards,
  title={Towards biologically plausible model-based reinforcement learning in recurrent spiking networks by dreaming new experiences},
  author={Capone, Cristiano and Paolucci, Pier Stanislao},
  journal={Scientific Reports},
  volume={14},
  number={1},
  pages={14656},
  year={2024},
  publisher={Nature Publishing Group UK London}
}

@article{deperrois2022learning,
  title={Learning cortical representations through perturbed and adversarial dreaming},
  author={Deperrois, Nicolas and Petrovici, Mihai A and Senn, Walter and Jordan, Jakob},
  journal={Elife},
  volume={11},
  pages={e76384},
  year={2022},
  publisher={eLife Sciences Publications Limited}
}

@article{kumar2023bayesian,
  title={Bayesian surprise predicts human event segmentation in story listening},
  author={Kumar, Manoj and Goldstein, Ariel and Michelmann, Sebastian and Zacks, Jeffrey M and Hasson, Uri and Norman, Kenneth A},
  journal={Cognitive science},
  volume={47},
  number={10},
  pages={e13343},
  year={2023},
  publisher={Wiley Online Library}
}

@article{stoianov2022hippocampal,
  title={The hippocampal formation as a hierarchical generative model supporting generative replay and continual learning},
  author={Stoianov, Ivilin and Maisto, Domenico and Pezzulo, Giovanni},
  journal={Progress in Neurobiology},
  volume={217},
  pages={102329},
  year={2022},
  publisher={Elsevier}
}

@article{baldassano2017discovering,
  title={Discovering event structure in continuous narrative perception and memory},
  author={Baldassano, Christopher and Chen, Janice and Zadbood, Asieh and Pillow, Jonathan W and Hasson, Uri and Norman, Kenneth A},
  journal={Neuron},
  volume={95},
  number={3},
  pages={709--721},
  year={2017},
  publisher={Elsevier}
}

@article{chang2022information,
  title={Information flow across the cortical timescale hierarchy during narrative construction},
  author={Chang, Claire HC and Nastase, Samuel A and Hasson, Uri},
  journal={Proceedings of the National Academy of Sciences},
  volume={119},
  number={51},
  pages={e2209307119},
  year={2022},
  publisher={National Academy of Sciences}
}

@article{lu2022neural,
  title={A neural network model of when to retrieve and encode episodic memories},
  author={Lu, Qihong and Hasson, Uri and Norman, Kenneth A},
  journal={elife},
  volume={11},
  pages={e74445},
  year={2022},
  publisher={eLife Sciences Publications Limited}
}

@article{hacker1980speed,
  title={Speed and accuracy of recency judgments for events in short-term memory.},
  author={Hacker, Michael J},
  journal={Journal of Experimental Psychology: Human Learning and Memory},
  volume={6},
  number={6},
  pages={651},
  year={1980},
  publisher={American Psychological Association}
}

@article{howard2018memory,
  title={Memory as perception of the past: compressed time in mind and brain},
  author={Howard, Marc W},
  journal={Trends in cognitive sciences},
  volume={22},
  number={2},
  pages={124--136},
  year={2018},
  publisher={Elsevier}
}

@article{addis2009constructive,
  title={Constructive episodic simulation of the future and the past: Distinct subsystems of a core brain network mediate imagining and remembering},
  author={Addis, Donna Rose and Pan, Ling and Vu, Mai-Anh and Laiser, Noa and Schacter, Daniel L},
  journal={Neuropsychologia},
  volume={47},
  number={11},
  pages={2222--2238},
  year={2009},
  publisher={Elsevier}
}

@article{hemmer2009integrating,
  title={Integrating episodic memories and prior knowledge at multiple levels of abstraction},
  author={Hemmer, Pernille and Steyvers, Mark},
  journal={Psychonomic bulletin \& review},
  volume={16},
  number={1},
  pages={80--87},
  year={2009},
  publisher={Springer}
}

@article{jimenez2024hipporag,
  title={Hipporag: Neurobiologically inspired long-term memory for large language models},
  author={Jimenez Gutierrez, Bernal and Shu, Yiheng and Gu, Yu and Yasunaga, Michihiro and Su, Yu},
  journal={Advances in Neural Information Processing Systems},
  volume={37},
  pages={59532--59569},
  year={2024}
}

@article{schacter2011memory,
  title={Memory distortion: An adaptive perspective},
  author={Schacter, Daniel L and Guerin, Scott A and Jacques, Peggy L St},
  journal={Trends in cognitive sciences},
  volume={15},
  number={10},
  pages={467--474},
  year={2011},
  publisher={Elsevier}
}

@article{hemmer2009bayesian,
  title={A Bayesian account of reconstructive memory},
  author={Hemmer, Pernille and Steyvers, Mark},
  journal={Topics in cognitive science},
  volume={1},
  number={1},
  pages={189--202},
  year={2009},
  publisher={Wiley Online Library}
}

@article{winocur2011memory,
  title={Memory transformation and systems consolidation},
  author={Winocur, Gordon and Moscovitch, Morris},
  journal={Journal of the International Neuropsychological Society},
  volume={17},
  number={5},
  pages={766--780},
  year={2011},
  publisher={Cambridge University Press}
}

@article{kahana1996associative,
  title={Associative retrieval processes in free recall},
  author={Kahana, Michael J},
  journal={Memory \& cognition},
  volume={24},
  number={1},
  pages={103--109},
  year={1996},
  publisher={Springer}
}

@article{moscovitch2005functional,
  title={Functional neuroanatomy of remote episodic, semantic and spatial memory: a unified account based on multiple trace theory},
  author={Moscovitch, Morris and Rosenbaum, R Shayna and Gilboa, Asaf and Addis, Donna Rose and Westmacott, Robyn and Grady, Cheryl and McAndrews, Mary Pat and Levine, Brian and Black, Sandra and Winocur, Gordon and others},
  journal={Journal of anatomy},
  volume={207},
  number={1},
  pages={35--66},
  year={2005},
  publisher={Wiley Online Library}
}

@article{nadel1997memory,
  title={Memory consolidation, retrograde amnesia and the hippocampal complex},
  author={Nadel, Lynn and Moscovitch, Morris},
  journal={Current opinion in neurobiology},
  volume={7},
  number={2},
  pages={217--227},
  year={1997},
  publisher={Elsevier}
}

@article{squire2015memory,
  title={Memory consolidation},
  author={Squire, Larry R and Genzel, Lisa and Wixted, John T and Morris, Richard G},
  journal={Cold Spring Harbor perspectives in biology},
  volume={7},
  number={8},
  pages={a021766},
  year={2015},
  publisher={Cold Spring Harbor Lab}
}

@article{squire1992memory,
  title={Memory and the hippocampus: a synthesis from findings with rats, monkeys, and humans.},
  author={Squire, Larry R},
  journal={Psychological review},
  volume={99},
  number={2},
  pages={195},
  year={1992},
  publisher={American Psychological Association}
}

@article{nicholas2026episodic,
  title={Episodic memory facilitates flexible decision-making via access to detailed events},
  author={Nicholas, Jonathan and Mattar, Marcelo G},
  journal={Nature Human Behaviour},
  pages={1--17},
  year={2026},
  publisher={Nature Publishing Group UK London}
}

@article{sederberg2011human,
  title={Human memory reconsolidation can be explained using the temporal context model},
  author={Sederberg, Per B and Gershman, Samuel J and Polyn, Sean M and Norman, Kenneth A},
  journal={Psychonomic bulletin \& review},
  volume={18},
  number={3},
  pages={455--468},
  year={2011},
  publisher={Springer}
}

@article{mason2025explicit,
  title={Explicit memory representations in decisions from experience},
  author={Mason, Alice and Lindskog, Marcus and Hertwig, Ralph and Wulff, Dirk U},
  journal={bioRxiv},
  pages={2025--10},
  year={2025},
  publisher={Cold Spring Harbor Laboratory}
}

@article{bornstein2017reminders,
  title={Reminders of past choices bias decisions for reward in humans},
  author={Bornstein, Aaron M and Khaw, Mel W and Shohamy, Daphna and Daw, Nathaniel D},
  journal={Nature Communications},
  volume={8},
  number={1},
  pages={15958},
  year={2017},
  publisher={Nature Publishing Group UK London}
}

@article{zhou2024episodic,
  title={Episodic retrieval for model-based evaluation in sequential decision tasks.},
  author={Zhou, Corey Y and Talmi, Deborah and Daw, Nathaniel D and Mattar, Marcelo G},
  journal={Psychological Review},
  year={2024},
  publisher={American Psychological Association}
}

@article{rosch1976structural,
  title={Structural bases of typicality effects.},
  author={Rosch, Eleanor and Simpson, Carol and Miller, R Scott},
  journal={Journal of Experimental Psychology: Human perception and performance},
  volume={2},
  number={4},
  pages={491},
  year={1976},
  publisher={American Psychological Association}
}

@article{mccloskey1980stimulus,
  title={The stimulus familiarity problem in semantic memory research},
  author={McCloskey, Michael},
  journal={Journal of Verbal Learning and Verbal Behavior},
  volume={19},
  number={4},
  pages={485--502},
  year={1980},
  publisher={Elsevier}
}

@article{addis2010episodic,
  title={Episodic simulation of past and future events in older adults: Evidence from an experimental recombination task.},
  author={Addis, Donna Rose and Musicaro, Regina and Pan, Ling and Schacter, Daniel L},
  journal={Psychology and aging},
  volume={25},
  number={2},
  pages={369},
  year={2010},
  publisher={American Psychological Association}
}

@article{schacter2015episodic,
  title={Episodic future thinking and episodic counterfactual thinking: Intersections between memory and decisions},
  author={Schacter, Daniel L and Benoit, Roland G and De Brigard, Felipe and Szpunar, Karl K},
  journal={Neurobiology of learning and memory},
  volume={117},
  pages={14--21},
  year={2015},
  publisher={Elsevier}
}

@article{shiffrin1997model,
  title={A model for recognition memory: REM—retrieving effectively from memory},
  author={Shiffrin, Richard M and Steyvers, Mark},
  journal={Psychonomic bulletin \& review},
  volume={4},
  number={2},
  pages={145--166},
  year={1997},
  publisher={Springer}
}

@article{gillund1984retrieval,
  title={A retrieval model for both recognition and recall.},
  author={Gillund, Gary and Shiffrin, Richard M},
  journal={Psychological review},
  volume={91},
  number={1},
  pages={1},
  year={1984},
  publisher={American Psychological Association}
}

@article{raaijmakers1981search,
  title={Search of associative memory.},
  author={Raaijmakers, Jeroen G and Shiffrin, Richard M},
  journal={Psychological review},
  volume={88},
  number={2},
  pages={93},
  year={1981},
  publisher={American Psychological Association}
}

@article{sims2018efficient,
  title={Efficient coding explains the universal law of generalization in human perception},
  author={Sims, Chris R},
  journal={Science},
  volume={360},
  number={6389},
  pages={652--656},
  year={2018},
  publisher={American Association for the Advancement of Science}
}

@article{anderson1989human,
  title={Human memory: An adaptive perspective.},
  author={Anderson, John R and Milson, Robert},
  journal={Psychological Review},
  volume={96},
  number={4},
  pages={703},
  year={1989},
  publisher={American Psychological Association}
}

@article{zenon2019information,
  title={An information-theoretic perspective on the costs of cognition},
  author={Zenon, Alexandre and Solopchuk, Oleg and Pezzulo, Giovanni},
  journal={Neuropsychologia},
  volume={123},
  pages={5--18},
  year={2019},
  publisher={Elsevier}
}

@article{marois2005capacity,
  title={Capacity limits of information processing in the brain},
  author={Marois, Ren{\'e} and Ivanoff, Jason},
  journal={Trends in cognitive sciences},
  volume={9},
  number={6},
  pages={296--305},
  year={2005},
  publisher={Elsevier}
}

@article{gershman2021rational,
  title={The rational analysis of memory},
  author={Gershman, Samuel J},
  journal={Oxford handbook of human memory.},
  year={2021},
  publisher={Oxford University Press Oxford, UK}
}

@article{sims2003implications,
  title={Implications of rational inattention},
  author={Sims, Christopher A},
  journal={Journal of monetary Economics},
  volume={50},
  number={3},
  pages={665--690},
  year={2003},
  publisher={Elsevier}
}

@article{postman1965short,
  title={Short-term temporal changes in free recall},
  author={Postman, Leo and Phillips, Laura W},
  journal={Quarterly journal of experimental psychology},
  volume={17},
  number={2},
  pages={132--138},
  year={1965},
  publisher={Taylor \& Francis}
}

@article{sederberg2008context,
  title={A context-based theory of recency and contiguity in free recall.},
  author={Sederberg, Per B and Howard, Marc W and Kahana, Michael J},
  journal={Psychological review},
  volume={115},
  number={4},
  pages={893},
  year={2008},
  publisher={American Psychological Association}
}

@article{farrell2002endogenous,
  title={An endogenous distributed model of ordering in serial recall},
  author={Farrell, Simon and Lewandowsky, Stephan},
  journal={Psychonomic bulletin \& review},
  volume={9},
  number={1},
  pages={59--79},
  year={2002},
  publisher={Springer}
}

@article{murdock1962serial,
  title={The serial position effect of free recall.},
  author={Murdock Jr, Bennet B},
  journal={Journal of experimental psychology},
  volume={64},
  number={5},
  pages={482},
  year={1962},
  publisher={American Psychological Association}
}

@article{suddendorf2007evolution,
  title={The evolution of foresight: What is mental time travel, and is it unique to humans?},
  author={Suddendorf, Thomas and Corballis, Michael C},
  journal={Behavioral and brain sciences},
  volume={30},
  number={3},
  pages={299--313},
  year={2007},
  publisher={Cambridge University Press}
}

@article{schacter2007cognitive,
  title={The cognitive neuroscience of constructive memory: Remembering the past and imagining the future},
  author={Schacter, Daniel L and Addis, Donna Rose},
  journal={Philosophical Transactions of the Royal Society B: Biological Sciences},
  volume={362},
  number={1481},
  pages={773--786},
  year={2007},
  publisher={The Royal Society London}
}

@article{schacter2017episodic,
  title={Episodic future thinking: Mechanisms and functions},
  author={Schacter, Daniel L and Benoit, Roland G and Szpunar, Karl K},
  journal={Current opinion in behavioral sciences},
  volume={17},
  pages={41--50},
  year={2017},
  publisher={Elsevier}
}

@article{jones2026,
  title={The Hippocampal Latent Diffusion Engine: A Computational Framework for Memory, Perception, and Cognitive Dysfunction},
  author={Jones, David and Breakspear, Michael},
  year={2026},
  journal={OSF}
}

@article{carr2011hippocampal,
  title={Hippocampal replay in the awake state: a potential substrate for memory consolidation and retrieval},
  author={Carr, Margaret F and Jadhav, Shantanu P and Frank, Loren M},
  journal={Nature neuroscience},
  volume={14},
  number={2},
  pages={147--153},
  year={2011},
  publisher={Nature Publishing Group US New York}
}

@article{van2024hierarchical,
  title={A hierarchical active inference model of spatial alternation tasks and the hippocampal-prefrontal circuit},
  author={Van de Maele, Toon and Dhoedt, Bart and Verbelen, Tim and Pezzulo, Giovanni},
  journal={Nature Communications},
  volume={15},
  number={1},
  pages={9892},
  year={2024},
  publisher={Nature Publishing Group UK London}
}

@article{bottini2020knowledge,
  title={Knowledge across reference frames: Cognitive maps and image spaces},
  author={Bottini, Roberto and Doeller, Christian F},
  journal={Trends in Cognitive Sciences},
  volume={24},
  number={8},
  pages={606--619},
  year={2020},
  publisher={Elsevier}
}

@article{george2021clone,
  title={Clone-structured graph representations enable flexible learning and vicarious evaluation of cognitive maps},
  author={George, Dileep and Rikhye, Rajeev V and Gothoskar, Nishad and Guntupalli, J Swaroop and Dedieu, Antoine and L{\'a}zaro-Gredilla, Miguel},
  journal={Nature communications},
  volume={12},
  number={1},
  pages={2392},
  year={2021},
  publisher={Nature Publishing Group UK London}
}

@article{behrens2018cognitive,
  title={What is a cognitive map? Organizing knowledge for flexible behavior},
  author={Behrens, Timothy EJ and Muller, Timothy H and Whittington, James CR and Mark, Shirley and Baram, Alon B and Stachenfeld, Kimberly L and Kurth-Nelson, Zeb},
  journal={Neuron},
  volume={100},
  number={2},
  pages={490--509},
  year={2018},
  publisher={Elsevier}
}

@article{hills2015exploration,
  title={Exploration versus exploitation in space, mind, and society},
  author={Hills, Thomas T and Todd, Peter M and Lazer, David and Redish, A David and Couzin, Iain D},
  journal={Trends in cognitive sciences},
  volume={19},
  number={1},
  pages={46--54},
  year={2015},
  publisher={Elsevier}
}

@article{lin2025neural,
  title={Neural sampling from cognitive maps supports goal-directed planning and imagination},
  author={Lin, Hui and Yang, Yukun and Zhao, Rong and Pezzulo, Giovanni and Maass, Wolfgang},
  journal={bioRxiv},
  pages={2025--05},
  year={2025},
  publisher={Cold Spring Harbor Laboratory}
}

@article{hobson2012waking,
  title={Waking and dreaming consciousness: neurobiological and functional considerations},
  author={Hobson, J Allan and Friston, Karl J},
  journal={Progress in neurobiology},
  volume={98},
  number={1},
  pages={82--98},
  year={2012},
  publisher={Elsevier}
}

@article{redish2016vicarious,
  title={Vicarious trial and error},
  author={Redish, A David},
  journal={Nature Reviews Neuroscience},
  volume={17},
  number={3},
  pages={147--159},
  year={2016},
  publisher={Nature Publishing Group UK London}
}

@article{buckner2010role,
  title={The role of the hippocampus in prediction and imagination},
  author={Buckner, Randy L},
  journal={Annual review of psychology},
  volume={61},
  number={1},
  pages={27--48},
  year={2010},
  publisher={Annual Reviews}
}

@article{mattar2018prioritized,
  title={Prioritized memory access explains planning and hippocampal replay},
  author={Mattar, Marcelo G. and Daw, Nathaniel D.},
  journal={Nature Neuroscience},
  volume={21},
  number={11},
  pages={1609--1617},
  year={2018},
  publisher={Nature Publishing Group}
}

@article{tompary2021semantic,
  title={Semantic influences on episodic memory distortions.},
  author={Tompary, Alexa and Thompson-Schill, Sharon L},
  journal={Journal of Experimental Psychology: General},
  volume={150},
  number={9},
  pages={1800},
  year={2021},
  publisher={American Psychological Association}
}

@article{sims2016rate,
  title={Rate--distortion theory and human perception},
  author={Sims, Chris R},
  journal={Cognition},
  volume={152},
  pages={181--198},
  year={2016},
  publisher={Elsevier}
}

@article{patterson2007you,
  title={Where do you know what you know? The representation of semantic knowledge in the human brain},
  author={Patterson, Karalyn and Nestor, Peter J and Rogers, Timothy T},
  journal={Nature reviews neuroscience},
  volume={8},
  number={12},
  pages={976--987},
  year={2007},
  publisher={Nature Publishing Group UK London}
}

@article{liu2019human,
  title={Human replay spontaneously reorganizes experience},
  author={Liu, Yunzhe and Dolan, Raymond J and Kurth-Nelson, Zeb and Behrens, Timothy EJ},
  journal={Cell},
  volume={178},
  number={3},
  pages={640--652},
  year={2019},
  publisher={Elsevier}
}

@inproceedings{he2016deep,
  title={Deep residual learning for image recognition},
  author={He, Kaiming and Zhang, Xiangyu and Ren, Shaoqing and Sun, Jian},
  booktitle={Proceedings of the IEEE conference on computer vision and pattern recognition},
  pages={770--778},
  year={2016}
}

@article{olafsdottir2018role,
  title={The role of hippocampal replay in memory and planning},
  author={{\'O}lafsd{\'o}ttir, H Freyja and Bush, Daniel and Barry, Caswell},
  journal={Current Biology},
  volume={28},
  number={1},
  pages={R37--R50},
  year={2018},
  publisher={Elsevier}
}

@article{pezzulo2017internally,
  title={Internally generated hippocampal sequences as a vantage point to probe future-oriented cognition},
  author={Pezzulo, Giovanni and Kemere, Caleb and Van Der Meer, Matthijs AA},
  journal={Annals of the New York Academy of Sciences},
  volume={1396},
  number={1},
  pages={144--165},
  year={2017},
  publisher={Wiley Online Library}
}

@article{schuck2019sequential,
  title={Sequential replay of nonspatial task states in the human hippocampus},
  author={Schuck, Nicolas W and Niv, Yael},
  journal={Science},
  volume={364},
  number={6447},
  pages={eaaw5181},
  year={2019},
  publisher={American Association for the Advancement of Science}
}

@article{pezzulo2021secret,
  title={The secret life of predictive brains: what’s spontaneous activity for?},
  author={Pezzulo, Giovanni and Zorzi, Marco and Corbetta, Maurizio},
  journal={Trends in cognitive sciences},
  volume={25},
  number={9},
  pages={730--743},
  year={2021},
  publisher={Elsevier}
}

@article{skaggs1996replay,
  title={Replay of neuronal firing sequences in rat hippocampus during sleep following spatial experience},
  author={Skaggs, William E and McNaughton, Bruce L},
  journal={Science},
  volume={271},
  number={5257},
  pages={1870--1873},
  year={1996},
  publisher={American Association for the Advancement of Science}
}

@article{pfeiffer2013hippocampal,
  title={Hippocampal place-cell sequences depict future paths to remembered goals},
  author={Pfeiffer, Brad E and Foster, David J},
  journal={Nature},
  volume={497},
  number={7447},
  pages={74--79},
  year={2013},
  publisher={Nature Publishing Group UK London}
}

@article{wimmer2020episodic,
  title={Episodic memory retrieval success is associated with rapid replay of episode content},
  author={Wimmer, G Elliott and Liu, Yunzhe and Vehar, Ne{\v{z}}a and Behrens, Timothy EJ and Dolan, Raymond J},
  journal={Nature neuroscience},
  volume={23},
  number={8},
  pages={1025--1033},
  year={2020},
  publisher={Nature Publishing Group US New York}
}

@article{norman2008computational,
  title={Computational models of episodic memory},
  author={Norman, Kenneth A and Detre, GJ and Polyn, Sean M},
  journal={The Cambridge handbook of computational psychology},
  volume={753},
  pages={189--225},
  year={2008},
  publisher={Cambridge University Press Cambridge, UK}
}

@article{norman2003modeling,
  title={Modeling hippocampal and neocortical contributions to recognition memory: a complementary-learning-systems approach.},
  author={Norman, Kenneth A and O'Reilly, Randall C},
  journal={Psychological review},
  volume={110},
  number={4},
  pages={611},
  year={2003},
  publisher={American Psychological Association}
}

@article{brady2008visual,
  title={Visual long-term memory has a massive storage capacity for object details},
  author={Brady, Timothy F and Konkle, Talia and Alvarez, George A and Oliva, Aude},
  journal={Proceedings of the National Academy of Sciences},
  volume={105},
  number={38},
  pages={14325--14329},
  year={2008},
  publisher={National Academy of Sciences}
}

@article{gershman2025key,
  title={Key-value memory in the brain},
  author={Gershman, Samuel J and Fiete, Ila and Irie, Kazuki},
  journal={Neuron},
  volume={113},
  number={11},
  pages={1694--1707},
  year={2025},
  publisher={Elsevier}
}

@article{spens2024generative,
  title={A generative model of memory construction and consolidation},
  author={Spens, Eleanor and Burgess, Neil},
  journal={Nature human behaviour},
  volume={8},
  number={3},
  pages={526--543},
  year={2024},
  publisher={Nature Publishing Group UK London}
}

@article{kahana2020computational,
  title={Computational models of memory search},
  author={Kahana, Michael J},
  journal={Annual Review of Psychology},
  volume={71},
  number={1},
  pages={107--138},
  year={2020},
  publisher={Annual Reviews}
}

@article{ngo2021contingency,
  title={Contingency of semantic generalization on episodic specificity varies across development},
  author={Ngo, Chi T and Benear, Susan L and Popal, Haroon and Olson, Ingrid R and Newcombe, Nora S},
  journal={Current Biology},
  volume={31},
  number={12},
  pages={2690--2697},
  year={2021},
  publisher={Elsevier}
}

@article{keresztes2018hippocampal,
  title={Hippocampal maturation drives memory from generalization to specificity},
  author={Keresztes, Attila and Ngo, Chi T and Lindenberger, Ulman and Werkle-Bergner, Markus and Newcombe, Nora S},
  journal={Trends in Cognitive Sciences},
  volume={22},
  number={8},
  pages={676--686},
  year={2018},
  publisher={Elsevier}
}

@article{kumaran2012generalization,
  title={Generalization through the recurrent interaction of episodic memories: a model of the hippocampal system.},
  author={Kumaran, Dharshan and McClelland, James L},
  journal={Psychological review},
  volume={119},
  number={3},
  pages={573},
  year={2012},
  publisher={American Psychological Association}
}

@article{nagy2020optimal,
  title={Optimal forgetting: Semantic compression of episodic memories},
  author={Nagy, David G and T{\"o}r{\"o}k, Bal{\'a}zs and Orb{\'a}n, Gerg{\H{o}}},
  journal={PLoS Computational Biology},
  volume={16},
  number={10},
  pages={e1008367},
  year={2020},
  publisher={Public Library of Science San Francisco, CA USA}
}

@article{fayyaz2022model,
  title={A model of semantic completion in generative episodic memory},
  author={Fayyaz, Zahra and Altamimi, Aya and Zoellner, Carina and Klein, Nicole and Wolf, Oliver T and Cheng, Sen and Wiskott, Laurenz},
  journal={Neural Computation},
  volume={34},
  number={9},
  pages={1841--1870},
  year={2022},
  publisher={MIT Press One Rogers Street, Cambridge, MA 02142-1209, USA journals-info~…}
}

@article{knowlton1995remembering,
  title={Remembering and knowing: two different expressions of declarative memory.},
  author={Knowlton, Barbara J and Squire, Larry R},
  journal={Journal of Experimental Psychology: Learning, Memory, and Cognition},
  volume={21},
  number={3},
  pages={699},
  year={1995},
  publisher={American Psychological Association}
}

@article{tulving1972episodic,
  title={Episodic and semantic memory},
  author={Tulving, Endel and others},
  journal={Organization of memory},
  volume={1},
  number={381-403},
  pages={1},
  year={1972},
  publisher={New York}
}

@article{tulving2002episodic,
  title={Episodic memory: From mind to brain},
  author={Tulving, Endel},
  journal={Annual review of psychology},
  volume={53},
  number={1},
  pages={1--25},
  year={2002},
  publisher={Annual Reviews 4139 El Camino Way, PO Box 10139, Palo Alto, CA 94303-0139, USA}
}

@incollection{atkinson1968human,
  title={Human memory: A proposed system and its control processes},
  author={Atkinson, Richard C and Shiffrin, Richard M},
  booktitle={Psychology of learning and motivation},
  volume={2},
  pages={89--195},
  year={1968},
  publisher={Elsevier}
}

@article{howard2002distributed,
  title={A distributed representation of temporal context},
  author={Howard, Marc W and Kahana, Michael J},
  journal={Journal of mathematical psychology},
  volume={46},
  number={3},
  pages={269--299},
  year={2002},
  publisher={Elsevier}
}

@article{kumaran2016learning,
  title={What learning systems do intelligent agents need? Complementary learning systems theory updated},
  author={Kumaran, Dharshan and Hassabis, Demis and McClelland, James L},
  journal={Trends in cognitive sciences},
  volume={20},
  number={7},
  pages={512--534},
  year={2016},
  publisher={Elsevier}
}

@article{schapiro2017complementary,
  title={Complementary learning systems within the hippocampus: a neural network modelling approach to reconciling episodic memory with statistical learning},
  author={Schapiro, Anna C and Turk-Browne, Nicholas B and Botvinick, Matthew M and Norman, Kenneth A},
  journal={Philosophical Transactions of the Royal Society B: Biological Sciences},
  volume={372},
  number={1711},
  pages={20160049},
  year={2017},
  publisher={The Royal Society}
}

@article{kingma2013auto,
  title={Auto-encoding variational bayes},
  author={Kingma, Diederik P and Welling, Max},
  journal={arXiv preprint arXiv:1312.6114},
  year={2013}
}

@article{lewis2020retrieval,
  title={Retrieval-augmented generation for knowledge-intensive nlp tasks},
  author={Lewis, Patrick and Perez, Ethan and Piktus, Aleksandra and Petroni, Fabio and Karpukhin, Vladimir and Goyal, Naman and K{\"u}ttler, Heinrich and Lewis, Mike and Yih, Wen-tau and Rockt{\"a}schel, Tim and others},
  journal={Advances in neural information processing systems},
  volume={33},
  pages={9459--9474},
  year={2020}
}

@article{teyler2007hippocampal,
  title={The hippocampal indexing theory and episodic memory: updating the index},
  author={Teyler, Timothy J and Rudy, Jerry W},
  journal={Hippocampus},
  volume={17},
  number={12},
  pages={1158--1169},
  year={2007},
  publisher={Wiley Online Library}
}

@article{teyler1986hippocampal,
  title={The hippocampal memory indexing theory.},
  author={Teyler, Timothy J and DiScenna, Pascal},
  journal={Behavioral neuroscience},
  volume={100},
  number={2},
  pages={147},
  year={1986},
  publisher={American Psychological Association}
}

@article{zacks2020event,
  title={Event perception and memory},
  author={Zacks, Jeffrey M},
  journal={Annual review of psychology},
  volume={71},
  number={1},
  pages={165--191},
  year={2020},
  publisher={Annual Reviews}
}

@article{mcclelland1995there,
  title={Why there are complementary learning systems in the hippocampus and neocortex: insights from the successes and failures of connectionist models of learning and memory.},
  author={McClelland, James L and McNaughton, Bruce L and O'Reilly, Randall C},
  journal={Psychological review},
  volume={102},
  number={3},
  pages={419},
  year={1995},
  publisher={American Psychological Association}
}

@article{burgess2018understanding,
  title={Understanding disentangling in beta-VAE},
  author={Burgess, Christopher P and Higgins, Irina and Pal, Arka and Matthey, Loic and Watters, Nick and Desjardins, Guillaume and Lerchner, Alexander},
  journal={arXiv preprint arXiv:1804.03599},
  year={2018}
}

@article{ramsauer2020hopfield,
  title={Hopfield networks is all you need},
  author={Ramsauer, Hubert and Sch{\"a}fl, Bernhard and Lehner, Johannes and Seidl, Philipp and Widrich, Michael and Adler, Thomas and Gruber, Lukas and Holzleitner, Markus and Pavlovi{\'c}, Milena and Sandve, Geir Kjetil and others},
  journal={arXiv preprint arXiv:2008.02217},
  year={2020}
}

@article{bhui2021resource,
  title={Resource-rational decision making},
  author={Bhui, Rahul and Lai, Lucy and Gershman, Samuel J},
  journal={Current Opinion in Behavioral Sciences},
  volume={41},
  pages={15--21},
  year={2021},
  publisher={Elsevier}
}

@article{pezzulo2024neural,
  title={Neural representation in active inference: Using generative models to interact with—and understand—the lived world},
  author={Pezzulo, Giovanni and D'Amato, Leo and Mannella, Francesco and Priorelli, Matteo and Van de Maele, Toon and Stoianov, Ivilin Peev and Friston, Karl},
  journal={Annals of the New York Academy of Sciences},
  volume={1534},
  number={1},
  pages={45--68},
  year={2024},
  publisher={Wiley Online Library}
}

@article{jakob2023rate,
  title={Rate-distortion theory of neural coding and its implications for working memory},
  author={Jakob, Anthony MV and Gershman, Samuel J},
  journal={Elife},
  volume={12},
  pages={e79450},
  year={2023},
  publisher={eLife Sciences Publications Limited}
}

@article{li2023rapid,
  title={Rapid memory encoding in a recurrent network model with behavioral time scale synaptic plasticity},
  author={Li, Pan Ye and Roxin, Alex},
  journal={PLoS Computational Biology},
  volume={19},
  number={8},
  pages={e1011139},
  year={2023},
  publisher={Public Library of Science San Francisco, CA USA}
}

@article{nicholas2025proactive,
  title={Proactive and reactive construction of memory-based preferences},
  author={Nicholas, Jonathan and Daw, Nathaniel D and Shohamy, Daphna},
  journal={Nature communications},
  volume={16},
  number={1},
  pages={1618},
  year={2025},
  publisher={Nature Publishing Group UK London}
}

@article{tarder2026adaptive,
  title={Adaptive episodic memory: how multiple memory representations drive behavior in humans and nonhumans},
  author={Tarder-Stoll, Hannah and Sekeres, Melanie J and Levine, Brian and Moscovitch, Morris},
  journal={Physiological Reviews},
  volume={106},
  number={2},
  pages={841--889},
  year={2026},
  publisher={American Physiological Society Rockville, MD}
}

@article{gilboa2021no,
  title={No consolidation without representation: Correspondence between neural and psychological representations in recent and remote memory},
  author={Gilboa, Asaf and Moscovitch, Morris},
  journal={Neuron},
  volume={109},
  number={14},
  pages={2239--2255},
  year={2021},
  publisher={Elsevier}
}

@article{brodt2018fast,
  title={Fast track to the neocortex: A memory engram in the posterior parietal cortex},
  author={Brodt, S and Gais, S and Beck, J and Erb, M and Scheffler, K and Sch{\"o}nauer, M},
  journal={Science},
  volume={362},
  number={6418},
  pages={1045--1048},
  year={2018},
  publisher={American Association for the Advancement of Science}
}

@article{brodt2016rapid,
  title={Rapid and independent memory formation in the parietal cortex},
  author={Brodt, Svenja and P{\"o}hlchen, Dorothee and Flanagin, Virginia L and Glasauer, Stefan and Gais, Steffen and Sch{\"o}nauer, Monika},
  journal={Proceedings of the National Academy of Sciences},
  volume={113},
  number={46},
  pages={13251--13256},
  year={2016},
  publisher={National Academy of Sciences}
}

@article{schacter2007remembering,
  title={Remembering the past to imagine the future: the prospective brain},
  author={Schacter, Daniel L and Addis, Donna Rose and Buckner, Randy L},
  journal={Nature reviews neuroscience},
  volume={8},
  number={9},
  pages={657--661},
  year={2007},
  publisher={Nature Publishing Group UK London}
}

@article{rugg2013brain,
  title={Brain networks underlying episodic memory retrieval},
  author={Rugg, Michael D and Vilberg, Kaia L},
  journal={Current opinion in neurobiology},
  volume={23},
  number={2},
  pages={255--260},
  year={2013},
  publisher={Elsevier}
}

@article{addis2018episodic,
  title={Are episodic memories special? On the sameness of remembered and imagined event simulation},
  author={Addis, Donna Rose},
  journal={Journal of the Royal Society of New Zealand},
  volume={48},
  number={2-3},
  pages={64--88},
  year={2018},
  publisher={Taylor \& Francis}
}

@article{renoult2019knowing,
  title={From knowing to remembering: the semantic--episodic distinction},
  author={Renoult, Louis and Irish, Muireann and Moscovitch, Morris and Rugg, Michael D},
  journal={Trends in cognitive sciences},
  volume={23},
  number={12},
  pages={1041--1057},
  year={2019},
  publisher={Elsevier}
}

@article{binder2011neurobiology,
  title={The neurobiology of semantic memory},
  author={Binder, Jeffrey R and Desai, Rutvik H},
  journal={Trends in cognitive sciences},
  volume={15},
  number={11},
  pages={527--536},
  year={2011},
  publisher={Elsevier}
}

@incollection{tulving1995organization,
  author    = {Tulving, Endel},
  title     = {Organization of memory: Quo vadis},
  booktitle = {The cognitive neurosciences},
  editor    = {Gazzaniga, Michael S.},
  pages     = {839--847},
  publisher = {MIT Press},
  year      = {1995},
  address   = {Cambridge, MA}
}

@article{tulving1998episodic,
  title={Episodic and declarative memory: role of the hippocampus},
  author={Tulving, Endel and Markowitsch, Hans J},
  journal={Hippocampus},
  volume={8},
  number={3},
  pages={198--204},
  year={1998},
  publisher={Wiley Online Library}
}

@article{tulving2001episodic,
  title={Episodic memory and common sense: how far apart?},
  author={Tulving, Endel},
  journal={Philosophical Transactions of the Royal Society of London. Series B: Biological Sciences},
  volume={356},
  number={1413},
  pages={1505--1515},
  year={2001},
  publisher={The Royal Society}
}

@article{wu2025simple,
  title={A simple model for Behavioral Time Scale Synaptic Plasticity (BTSP) provides content addressable memory with binary synapses and one-shot learning},
  author={Wu, Yujie and Maass, Wolfgang},
  journal={Nature communications},
  volume={16},
  number={1},
  pages={342},
  year={2025},
  publisher={Nature Publishing Group UK London}
}

@article{nelson2013co,
  title={The co-evolution of knowledge and event memory.},
  author={Nelson, Angela B and Shiffrin, Richard M},
  journal={Psychological Review},
  volume={120},
  number={2},
  pages={356},
  year={2013},
  publisher={American Psychological Association}
}

@inproceedings{perez2018film,
  title={Film: Visual reasoning with a general conditioning layer},
  author={Perez, Ethan and Strub, Florian and De Vries, Harm and Dumoulin, Vincent and Courville, Aaron},
  booktitle={Proceedings of the AAAI conference on artificial intelligence},
  volume={32},
  number={1},
  year={2018}
}

@article{friedman2022vendi,
  title={The vendi score: A diversity evaluation metric for machine learning},
  author={Friedman, Dan and Dieng, Adji Bousso},
  journal={arXiv preprint arXiv:2210.02410},
  year={2022}
}

@article{hupbach2007reconsolidation,
  title={Reconsolidation of episodic memories: A subtle reminder triggers integration of new information},
  author={Hupbach, Almut and Gomez, Rebecca and Hardt, Oliver and Nadel, Lynn},
  journal={Learning \& memory},
  volume={14},
  number={1-2},
  pages={47--53},
  year={2007},
  publisher={Cold Spring Harbor Lab}
}

@article{zaki2025offline,
  title={Offline ensemble co-reactivation links memories across days},
  author={Zaki, Yosif and Pennington, Zachary T and Morales-Rodriguez, Denisse and Bacon, Madeline E and Ko, BumJin and Francisco, Taylor R and LaBanca, Alexa R and Sompolpong, Patlapa and Dong, Zhe and Lamsifer, Sophia and others},
  journal={Nature},
  volume={637},
  number={8044},
  pages={145--155},
  year={2025},
  publisher={Nature Publishing Group UK London}
}

@article{eichenbaum1999hippocampus,
  title={The hippocampus and mechanisms of declarative memory},
  author={Eichenbaum, Howard},
  journal={Behavioural brain research},
  volume={103},
  number={2},
  pages={123--133},
  year={1999},
  publisher={Elsevier}
}

@article{gershman2013neural,
  title={Neural context reinstatement predicts memory misattribution},
  author={Gershman, Samuel J and Schapiro, Anna C and Hupbach, Almut and Norman, Kenneth A},
  journal={Journal of Neuroscience},
  volume={33},
  number={20},
  pages={8590--8595},
  year={2013},
  publisher={Society for Neuroscience}
}

@article{roediger2011critical,
  title={The critical role of retrieval practice in long-term retention},
  author={Roediger, Henry L and Butler, Andrew C},
  journal={Trends in cognitive sciences},
  volume={15},
  number={1},
  pages={20--27},
  year={2011},
  publisher={Elsevier}
}

@article{sohn2015learning,
  title={Learning structured output representation using deep conditional generative models},
  author={Sohn, Kihyuk and Lee, Honglak and Yan, Xinchen},
  journal={Advances in neural information processing systems},
  volume={28},
  year={2015}
}

\clearpage
\appendix
\section*{Supplementary Material}
\setcounter{figure}{0}
\renewcommand{\thefigure}{S\arabic{figure}}

\subsection*{Distortions due to lossy compression of values in Episodic Memory}

\begin{figure}[H]
    \centering
    \includegraphics[width=0.9\textwidth]{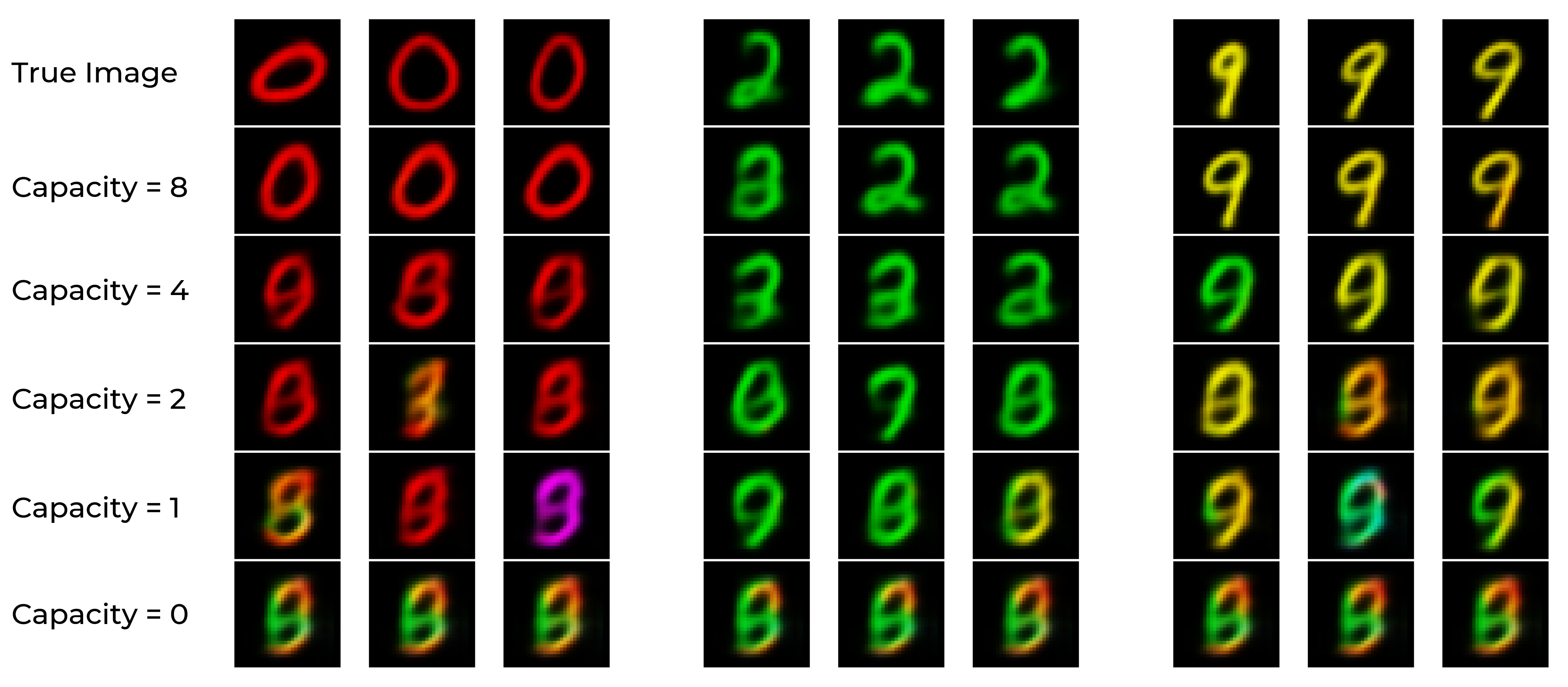}
    \caption{Image reconstructions at varying levels of item embedding compression. The embeddings are compressed by the Hippocampal-VAE at different capacity levels and subsequently decoded using a fixed Cortical-VAE. The top row shows the true images, and each following row corresponds to reconstructions from embeddings compressed at progressively different Hippocampal-VAE capacities.}
    \label{fig:hallucinations}
\end{figure}

In standard scenarios, image reconstruction was performed by passing the item embedding $h = [z, e_S, e_C]$ into the Cortical-VAE decoder, where $h$ encodes the value retrieved from the Episodic Memory RAG system. When this value is subject to lossy compression due to the limited capacity of the Hippocampal-VAE, a noisy version of the embedding is instead provided to the Cortical-VAE decoder. Figure \ref{fig:hallucinations} illustrates the reconstructed images under varying capacity constraints of the Hippocampal-VAE. The resulting image chimerization distorted both digit shapes and colors, as the conditional embeddings $e_S$ and $e_C$ were corrupted by noise, along with the latent state $z$.

\subsection*{Conditional $\beta$-VAE Training}

\begin{figure}[H]
    \centering
    \includegraphics[width=0.9\textwidth]{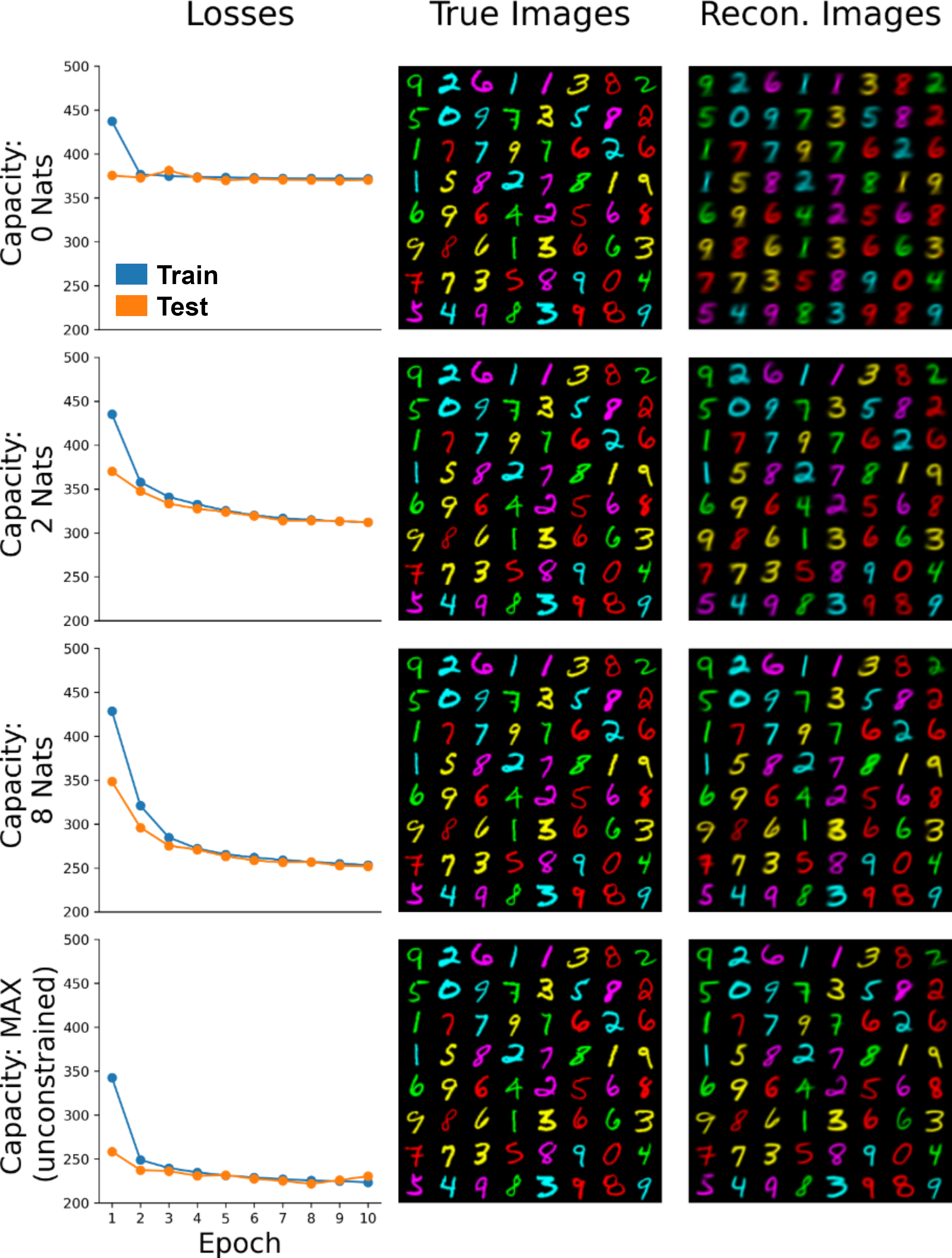}
    \caption{Reconstruction losses of conditional $\beta$-VAEs trained under different capacity constraints. A set of true images from the test set, and their reconstructions are shown in the second and third columns, respectively.}
    \label{fig:losses}
\end{figure}

All conditional $\beta$-VAEs used to implement the Cortical-VAE under different capacity constraints were trained with the same architecture. Training was run for 10 epochs, which corresponded to approximately 8,000 iterations per model and was sufficient for most models to converge (Figure \ref{fig:losses}). In total, we trained five capacity-constrained models with $C \in {0, 1, 2, 4, 8}$, as well as one unconstrained model. The unconstrained model—referred to in the main text as the max-capacity model—was optimized to achieve the best reconstruction performance among all Cortical-VAEs. Unlike the capacity-limited variants, it could generate novel images by sampling from the latent variable $z$, a capability inhibited in the constrained models due to restricted loss optimization. Whenever image simulation was required in an experiment, we used the unconstrained Cortical-VAE.

\end{document}